\documentclass[%
  reprint,
  amsmath,amssymb,
  aps,
  prd,
  floatfix,
  superscriptaddress,
  longbibliography
]{revtex4-2}

\usepackage[T1]{fontenc}
\usepackage[utf8]{inputenc}
\usepackage{lmodern}
\usepackage{graphicx}
\usepackage{bm}
\usepackage{dcolumn}
\usepackage[hidelinks]{hyperref}
\usepackage{siunitx}
\usepackage{xspace}
\usepackage{amsmath,amssymb,bm}
\usepackage{mathtools}
\usepackage{verbatim}
\usepackage{soul} 
\usepackage{xcolor}

\newcommand{\T}{\mathrm{T}}
\newcommand{\herm}{\dagger}

\begin{document}

\title{Scalar-Magnetometer Search for Ultralight Dark Photon Dark Matter\\
with a Single-Site, Two-Sensor Array:\\
A 6-Channel DTFT Likelihood Analysis with Scalar Optically Pumped Magnetometers}

\author{Peisen Zhao}
\email{ge47zis@mytum.de}
\affiliation{Department of Physics, Technical University of Munich, 85748 Garching, Germany}

\author{Ole Behrens}
\affiliation{Department of Physics, Technical University of Munich, 85748 Garching, Germany}

\author{Maja Benning}
\affiliation{Department of Physics, Technical University of Munich, 85748 Garching, Germany}

\author{Peter Fierlinger}
\affiliation{Department of Physics, Technical University of Munich, 85748 Garching, Germany}

\author{Xuefen Han}
\affiliation{Department of Physics, Technical University of Munich, 85748 Garching, Germany}
\affiliation{Heinz Maier-Leibnitz Zentrum, Technical University of Munich, Lichtenbergstraße 1, 85748 Garching, Germany}

\author{Maximilian Huber}
\affiliation{Department of Physics, Technical University of Munich, 85748 Garching, Germany}

\author{Florian Kuchler}
\affiliation{Department of Physics, Technical University of Munich, 85748 Garching, Germany}

\author{Yevgeny V. Stadnik}
\affiliation{School of Physics, The University of Sydney, Sydney, NSW 2006, Australia}

\author{Philipp Wunderl}
\affiliation{Department of Physics, Technical University of Munich, 85748 Garching, Germany}

\date{\today}


\begin{abstract}
We report a laboratory search for ultralight dark photon dark matter using a \emph{single-site}, \emph{two-sensor} array of commercial scalar optically pumped magnetometers (OPMs). In the low-frequency regime where the Earth--ionosphere system acts as an electromagnetic transducer, the expected magnetic signal is a narrow-band triplet of frequencies. This signature consists of a central peak at the dark photon's Compton frequency, accompanied by two sidebands shifted by Earth's sidereal rotation frequency. Because scalar magnetometers measure field magnitude, the observable signal is the projection of the oscillating dark photon magnetic field onto the direction of the large, local geomagnetic field. This preserves the crucial triplet signature in the resulting time series data. Analyzing 10.5 hours of continuous data, we construct a six-channel complex data vector by evaluating the discrete-time Fourier transform (DTFT) for both sensors directly at the three physical frequencies of the signal triplet. Assuming complex-Gaussian noise, we develop a likelihood framework to set robust, frequency-resolved upper limits on the kinetic-mixing parameter $\varepsilon$, which governs the coupling between Standard Model photons and dark photons. Within the mass range $4 \times 10^{-15}\,\text{eV} \lesssim m_{A'} \lesssim 3 \times 10^{-14}\,\text{eV}$, we obtain the most stringent direct laboratory limits to date on the kinetic-mixing parameter, which are complementary to existing astrophysical bounds, including those inferred from observations of the Leo~T dwarf galaxy.
\end{abstract}

\maketitle

\section{Introduction}
\label{sec:introduction}

\subsection{Ultralight Dark Matter and Dark Photons}
\label{subsec:intro_dark_photons}

The existence of dark matter (DM) is overwhelmingly supported by a vast body of astrophysical and cosmological observations, including galactic rotation curves, gravitational lensing, and the cosmic microwave background anisotropies \cite{Bertone2018_History, Zwicky1933, Rubin1980}. Despite constituting approximately 85\% of the universe's matter content, its fundamental nature remains one of the most profound mysteries in modern physics. While weakly interacting massive particles (WIMPs) have long been a primary focus of experimental searches, the lack of conclusive evidence has motivated a broader exploration of alternative candidates, particularly in the sub-eV mass range \cite{Graham2015_Review, Irastorza2018_Review, Antypas:2022asj}. 

In the ultralight regime ($m \ll \SI{1}{\electronvolt}$), bosonic DM candidates exhibit a macroscopic de Broglie wavelength. For a virialized galactic halo, this leads to an extremely high occupation number, allowing the DM field to be described as a classical wave oscillating at its Compton frequency, $\omega \approx m$ (in the natural units $\hbar = c = 1$) \cite{Hui2017_Ultralight}. This wavelike nature opens up a distinct class of detection strategies based on searching for coherent, time-varying signals rather than discrete particle scattering events.

Among the well-motivated ultralight candidates is the dark photon ($A'$), a hypothetical spin-1 vector boson arising from a new U(1) gauge symmetry associated with a "hidden sector" \cite{Okun1982, Holdom1986, Pospelov2008}. The primary portal for its interaction with the Standard Model is through kinetic mixing with the ordinary photon, described by the Lagrangian term
\begin{equation}
  \mathcal{L}_{\text{mix}} = -\frac{\varepsilon}{2} F_{\mu\nu} F'^{\mu\nu},
  \label{eq:kinetic_mixing_lagrangian}
\end{equation}
where $F_{\mu\nu}$ and $F'^{\mu\nu}$ are the field strength tensors for the Standard Model photon and the dark photon, respectively, and $\varepsilon$ is the dimensionless kinetic mixing parameter. If the dark photon constitutes the local DM, this mixing induces an effective oscillating current that sources a real, albeit very weak, electromagnetic field that oscillates at the dark photon's Compton frequency \cite{FedderkeTransducer}. This makes the ambient dark photon field detectable through precision magnetometry.

\subsection{The Predicted Dark Photon Signal Signature}
\label{subsec:intro_transducer}

In the low-frequency regime, where the dark photon's Compton wavelength $\lambda_{A'}$ is much larger than Earth's radius $R$ (i.e., $m_{A'} R \ll 1$), Earth and its surrounding ionosphere act as a natural transducer, converting the pervasive Galactic dark photon field into a measurable magnetic signal \cite{FedderkeTransducer, FedderkeSuperMAG}. 
The highly conductive surface of Earth and the plasma of the ionosphere form a roughly spherical resonant cavity. The ambient dark photon field, acting as an effective current source, drives electromagnetic modes within this cavity. The boundary conditions imposed by these conductive layers amplify the resulting magnetic field $\mathbf{B}_a(t)$ in the atmosphere, making it potentially detectable by ground-based magnetometers.

The expected signal possesses a unique spectral signature defined by its extreme monochromaticity. Due to the low velocity dispersion of the virialized dark matter halo ($v_{\text{DM}} \sim 10^{-3}$), the signal is expected to be exceptionally narrow, corresponding to a quality factor of $Q \approx 1/v_{\text{DM}}^2 \sim 10^6$. This high Q-factor makes the signal highly distinguishable from broadband noise. The primary oscillation occurs at the dark photon's Compton frequency, $f_{A'} = m_{A'} / (2\pi)$. However, an Earth-based observatory is rotating within the Galactic DM halo, which is quasi-static in the galactic frame. This rotation induces a daily modulation in the measured amplitude and phase of the signal, as the orientation of the detector with respect to the dark photon field's polarization vector changes.

This modulation is governed by Earth's sidereal rotation frequency, $f_d \approx (\SI{86164.09}{\second})^{-1}$. In the frequency domain, this amplitude and phase modulation decomposes the single narrow-band signal into a distinct \textbf{triplet structure}, with power concentrated at the three frequencies:
\begin{equation}
  f \in \{f_{A'}, f_{A'} - f_d, f_{A'} + f_d\},
  \label{eq:frequency_triplet}
\end{equation}
assuming that $f_{A'} > f_d$. 
The central peak at $f_{A'}$ corresponds to the component of the DM field polarization aligned with Earth's axis of rotation, while the two sidebands at $f_{A'} \pm f_d$ arise from the components perpendicular to the rotation axis \cite{FedderkeSuperMAG}. This sharp, predictable triplet signature provides a powerful tool for distinguishing a potential dark photon signal from terrestrial and instrumental noise. The primary objective of this work is to search for this characteristic frequency triplet using a dedicated two-sensor scalar magnetometer array and a specialized likelihood analysis tailored to this signal morphology.

\subsection{Previous Work and Our Contribution}
\label{subsec:intro_contribution}

The search for such terrestrial signatures of dark photon DM has been pioneered by analyses of magnetometer data. Most notably, the work by Fedderke \textit{et al.} utilized the SuperMAG collaboration's global network of vector magnetometers to set competitive limits by exploiting the signal's predicted spatial coherence and vectorial nature across geographically dispersed stations \cite{FedderkeSuperMAG}. Other terrestrial searches, such as SNIPE Hunt \cite{SNIPE_Hunt_Placeholder} and AMAILS \cite{AMAILS_Placeholder}, have also contributed to constraining this parameter space. These large-scale or multi-sensor efforts are powerful in their use of existing geophysical infrastructure and their ability to probe spatial correlations.

Complementary to these approaches are dedicated experiments at carefully selected, low-noise locations. Such experiments, often employing state-of-the-art magnetometers, benefit from controlled environments, long coherent integration times, and thorough characterization of instrumental systematics. This strategy prioritizes sensitivity and noise control at a single location over the global correlation signatures accessible to large networks. Crucially, the high sensitivity of this approach allows for competitive constraints to be set with observation times that can be orders of magnitude shorter than those required by global network analyses. For instance, the results presented in this work are derived from only 10.5 hours of data, yet achieve a sensitivity comparable to, and in some frequency bands exceeding, limits derived from decades of archival data.

The present work carves a distinct path within this landscape, demonstrating the feasibility of a high-sensitivity search using a deliberately \emph{minimal} configuration: a single experimental site instrumented with two scalar optically pumped magnetometers (OPMs). While sacrificing the direct vectorial information and global coherence checks of a large network, our approach leverages the high intrinsic sensitivity of modern OPMs and a tailored analysis framework to achieve competitive bounds. Our key contributions are threefold:

\begin{enumerate}
    \item \textbf{Minimalist Experimental Configuration:} We demonstrate a powerful yet resource-efficient approach using only two scalar magnetometers at a single site. This setup significantly reduces the logistical complexity and systematics associated with inter-site calibration and timing synchronization, providing a scalable model for future dedicated searches.

    \item \textbf{Scalar Projection Signal Model:} For scalar devices that measure the field magnitude, the primary observable is the projection of the oscillating dark photon magnetic field, $\mathbf{B}_a(t)$, onto the direction of the large, local background magnetic field, $\mathbf{B}_b$. The resulting scalar signal is given by $s(t) = \hat{\mathbf{B}}_b \cdot \mathbf{B}_a(t)$, and we develop a complete theoretical model for it. We show that this projection fully preserves the crucial frequency triplet structure, which remains the cornerstone of the search. This work establishes a formal bridge between the vectorial dark photon signal and the scalar data streams.

    \item \textbf{Advanced DTFT Likelihood Framework:} We introduce a novel likelihood analysis based on a six-channel analysis vector, constructed by evaluating the discrete-time Fourier transform (DTFT) of the data streams directly at the three physical frequencies of the signal triplet. This method entirely circumvents the grid-mismatch systematics and spectral leakage modeling complexities inherent to Fast Fourier Transform (FFT)-based methods that rely on the nearest frequency bin. Our framework provides a statistically robust and computationally efficient means of extracting limits from the data.
\end{enumerate}
Together, these contributions establish a robust and reproducible framework for single-site scalar magnetometer searches for ultralight dark matter, opening a new avenue in the broader experimental program.

\subsection{Organization of this Paper}
\label{subsec:intro_organization}

The remainder of this paper is organized as follows. Section~\ref{sec:theory} provides a detailed derivation of the scalar-projected signal model and the expected frequency triplet structure. Section~\ref{sec:instrumentation} describes our experimental setup, including the optically pumped magnetometers, the low-noise site, and on-site calibration procedures. Section~\ref{sec:data_preproc} outlines the data preprocessing pipeline, from raw timestamps to cleaned, prepared time series, and defines the construction of the analysis vector via the DTFT. In Sec.~\ref{sec:analysis}, we construct the six-channel likelihood framework, detailing the model for the mean vector and the empirical estimation of the noise covariance matrix. Our main results, the derived upper limits on the kinetic-mixing parameter $\varepsilon$, are presented and compared with existing constraints in Sec.~\ref{sec:results}. A thorough discussion of systematic uncertainties and the results of our signal injection validation tests is provided in Sec.~\ref{sec:systematics}. Finally, we summarize our findings and discuss future prospects in Sec.~\ref{sec:conclusion}. Technical details on the DTFT implementation and covariance estimation are provided in Appx.~\ref{app:dft_details} and \ref{app:covariance_estimation}.

\section{Theoretical Framework}
\label{sec:theory}

The theoretical basis for this search connects the global properties of the dark photon DM field to a local, measurable time series from a scalar magnetometer. This section details the signal model for the induced magnetic field and derives the specific observable that forms the basis of our likelihood analysis.

\subsection{Dark Photon Signal Model}
\label{subsec:transducer_signal}

In the low-frequency regime where the dark photon mass $m_{A'}$ is small such that its Compton wavelength greatly exceeds Earth's radius $R$ ($m_{A'} R \ll 1$), the Earth--ionosphere system acts as a coherent transducer \cite{FedderkeTransducer}. The ambient dark photon field, oscillating at its Compton frequency $f_{A'} \approx m_{A'}/(2\pi)$, sources an effective current which drives electromagnetic modes within the terrestrial cavity. As derived in Ref.~\cite{FedderkeSuperMAG}, the resulting magnetic field $\mathbf{B}_{a}(t)$ (we use the subscript `$a$' to denote the signal field hereafter) in the Earth-fixed frame at a geographic location $\Omega$ is given by
\begin{equation}
\begin{split}
\mathbf{B}_{a}(\Omega,t) ={}& \sqrt{\frac{4\pi}{3}}\,
\frac{m_{A'}R}{\,2-(m_{A'}R)^2\,}\;
\varepsilon\,m_{A'}\; \\
& \times \mathrm{Re}\!\left[
\sum_{m=-1}^{+1}
A'_m\,
\boldsymbol{\Phi}_{1m}(\Omega)\;
e^{-\,2\pi i\,\left(f_{A'}-m f_d\right)\,t}
\right].
\end{split}
\label{eq:Ba_vector_signal}
\end{equation}
This expression encapsulates the key features of the expected signal. The overall amplitude is proportional to the kinetic mixing parameter $\varepsilon$, which we aim to constrain. The term $A'_m$ represents the complex amplitude of the dark photon field's polarization state in the inertial (Galactic) frame, with $m=0$ corresponding to polarization along Earth's rotation axis, and $m=\pm 1$ corresponding to the two orthogonal transverse polarizations. These amplitudes are stochastic variables, assumed to be constant over the dark matter coherence time, $\tau_c \sim (f_{A'} v_{\text{DM}}^2)^{-1}$, where $v_{\text{DM}} \sim 10^{-3}$ is the virial velocity of the DM halo locally.

The vectorial and geometric properties of the signal are captured by the $\ell=1$ toroidal vector spherical harmonics (VSH), $\boldsymbol{\Phi}_{1m}(\Omega)$. These functions provide the mapping from the global polarization state in the inertial frame to the local magnetic field vector at a specific point on Earth's surface. Finally, the exponential term reveals the characteristic frequency triplet structure, where each polarization component $m$ contributes power at a distinct frequency $f_{A'} - m f_d$, modulated by Earth's sidereal rotation frequency $f_d$.

\subsection{Scalar Magnetometer Response: The Projected Signal}
\label{subsec:scalar_response}

The geomagnetic field strength at the experimental site is $|\mathbf{B}_b| \approx \SI{50}{\micro\tesla}$. In contrast, the dark photon signal $\mathbf{B}_a(t)$ is sought at a scale set by the sensitivity of our magnetometers, which corresponds to the picotesla-to-femtotesla range. Given this extreme separation of scales, where $|\mathbf{B}_a|/|\mathbf{B}_b| \lesssim 10^{-8}$, we can perform a first-order Taylor expansion of the magnitude around $\mathbf{B}_a = \mathbf{0}$:
\begin{equation}
  \mathbf{B}_{\text{total}}(t) = \mathbf{B}_b + \mathbf{B}_a(t).
\end{equation}
The sensor's output, $B(t)$, is the norm of this total field, $B(t) = \lVert \mathbf{B}_b + \mathbf{B}_a(t) \rVert$.

Given the extreme weakness of the expected dark matter signal relative to the geomagnetic field ($|\mathbf{B}_a| \ll |\mathbf{B}_b|$), we can perform a first-order Taylor expansion of the magnitude around $\mathbf{B}_a = \mathbf{0}$:
\begin{align}
  B(t) &= \sqrt{(\mathbf{B}_b + \mathbf{B}_a) \cdot (\mathbf{B}_b + \mathbf{B}_a)} \nonumber \\
       &= \sqrt{|\mathbf{B}_b|^2 + 2\mathbf{B}_b \cdot \mathbf{B}_a + |\mathbf{B}_a|^2} \nonumber \\
       &= |\mathbf{B}_b| \sqrt{1 + \frac{2\mathbf{B}_b \cdot \mathbf{B}_a}{|\mathbf{B}_b|^2} + \frac{|\mathbf{B}_a|^2}{|\mathbf{B}_b|^2}} \nonumber \\
       &\approx |\mathbf{B}_b| \left(1 + \frac{\mathbf{B}_b \cdot \mathbf{B}_a}{|\mathbf{B}_b|^2}\right).
       \label{eq:taylor_expansion}
\end{align}
The final approximation in Eq.~\eqref{eq:taylor_expansion} is achieved by performing a Taylor expansion of the square root to first order in the small quantity $|\mathbf{B}_a|/|\mathbf{B}_b|$, neglecting all terms of order $\mathcal{O}((|\mathbf{B}_a|/|\mathbf{B}_b|)^2)$ and higher. This yields the following crucial linearization for the scalar magnetometer signal:
\begin{equation}
  B(t) \approx |\mathbf{B}_b| + \frac{\mathbf{B}_b}{|\mathbf{B}_b|} \cdot \mathbf{B}_a(t) = |\mathbf{B}_b| + \hat{\mathbf{B}}_b \cdot \mathbf{B}_a(t),
  \label{eq:scalar_linearization}
\end{equation}
where $\hat{\mathbf{B}}_b$ is the unit vector in the direction of the local background magnetic field.

The first term, $|\mathbf{B}_b|$, is a large, nearly constant DC offset on the order of tens of microteslas. This term, along with other slow drifts from diurnal variations or instrumental effects, contains no information about the high-frequency dark photon signal and is removed during the data preprocessing stage (see Sec.~\ref{sec:data_preproc}). The remaining time-varying component,
\begin{equation}
  s(t) \equiv \hat{\mathbf{B}}_b \cdot \mathbf{B}_a(t),
  \label{eq:s_projection_def}
\end{equation}
is the primary observable for our analysis. It represents the projection of the vectorial dark photon magnetic field onto the direction of the local geomagnetic field. Crucially, because this projection is a linear operation, the resulting scalar time series $s(t)$ fully preserves the frequency triplet structure inherent to the vector field $\mathbf{B}_a(t)$. Our analysis is therefore designed to search for this signature within the projected data streams from our two magnetometers.

\subsection{Formation of the Frequency Triplet}
\label{subsec:theory_triplet}

To explicitly derive the frequency content of the projected scalar signal $s(t)$, we express both the background field direction $\hat{\mathbf{B}}_b$ and the dark photon signal $\mathbf{B}_a(t)$ in a local, Earth-fixed orthonormal basis $(\hat{r}, \hat{\theta}, \hat{\phi})$. This basis is derived from a standard geocentric spherical coordinate system where the colatitude $\theta$ is measured from the North Pole and the longitude $\phi$ increases eastward. Consequently, at the observer's location, the basis vectors are defined as follows: $\hat{r}$ points radially outward (local zenith), $\hat{\theta}$ points in the direction of increasing colatitude (south), and $\hat{\phi}$ points in the direction of increasing longitude (east). This right-handed coordinate system forms the reference for our signal projection model.

At the observation site, the unit vector of the background field can be decomposed as:
\begin{equation}
  \hat{\mathbf{B}}_b = u_r \hat{r} + u_\theta \hat{\theta} + u_\phi \hat{\phi},
  \label{eq:bb_local_basis}
\end{equation}
where $(u_r, u_\theta, u_\phi)$ are the direction cosines determined from on-site calibration (see Sec.~\ref{sec:instrumentation}). The transducer solution for the dark photon field, Eq.~\eqref{eq:Ba_vector_signal}, is based on toroidal vector spherical harmonics, which are purely tangential ($\boldsymbol{\Phi}_{1m} \cdot \hat{r} = 0$). Consequently, the induced magnetic field $\mathbf{B}_a(t)$ near Earth's surface has no radial component,
\begin{equation}
  \mathbf{B}_a(\Omega,t) = B_\theta(\Omega,t)\,\hat{\theta} + B_\phi(\Omega,t)\,\hat{\phi}.
\end{equation}
The projected signal $s(t) = \hat{\mathbf{B}}_b \cdot \mathbf{B}_a(t)$ therefore simplifies to a linear combination of only the tangential components:
\begin{equation}
  s(t) = u_\theta B_\theta(\Omega,t) + u_\phi B_\phi(\Omega,t).
  \label{eq:s_projection_tangential}
\end{equation}
Since both $B_\theta(t)$ and $B_\phi(t)$ contain the superposition of the three frequency components from Eq.~\eqref{eq:Ba_vector_signal}, the scalar signal $s(t)$ naturally inherits this structure. By substituting the harmonic expansion of the vector components, we can express $s(t)$ as a sum of three distinct oscillations:
\begin{equation}
\label{eq:s_three_lines}
\begin{aligned}
  s(t) ={}& \mathrm{Re}\left\{ S_0(\Omega)\,e^{-i2\pi f_{A'} t} \right\} \\
  &+ \mathrm{Re}\left\{ S_{-}(\Omega)\,e^{-i2\pi (f_{A'}-f_d) t} \right\} \\
  &+ \mathrm{Re}\left\{ S_{+}(\Omega)\,e^{-i2\pi (f_{A'}+f_d) t} \right\},
\end{aligned}
\end{equation}
where $S_0$, $S_-$, and $S_+$ are the complex amplitudes of the central peak and the two sidebands, respectively. These amplitudes encapsulate the full dependence on the site geometry and the DM field's polarization.

They are defined by collecting terms from Eq.~\eqref{eq:Ba_vector_signal} for each value of $m$:
\begin{equation}
\label{eq:S_complex_amplitudes}
\begin{aligned}
  S_0(\Omega)   &= \mathcal{A}_{\text{phys}}\, \varepsilon\, A'_0   \left[ u_\theta (\boldsymbol{\Phi}_{1,0}\cdot\hat{\theta}) + u_\phi (\boldsymbol{\Phi}_{1,0}\cdot\hat{\phi}) \right], \\
  S_-(\Omega)   &= \mathcal{A}_{\text{phys}}\, \varepsilon\, A'_{+1} \left[ u_\theta (\boldsymbol{\Phi}_{1,+1}\cdot\hat{\theta}) + u_\phi (\boldsymbol{\Phi}_{1,+1}\cdot\hat{\phi}) \right], \\
  S_+(\Omega)   &= \mathcal{A}_{\text{phys}}\, \varepsilon\, A'_{-1} \left[ u_\theta (\boldsymbol{\Phi}_{1,-1}\cdot\hat{\theta}) + u_\phi (\boldsymbol{\Phi}_{1,-1}\cdot\hat{\phi}) \right].
\end{aligned}
\end{equation}
Here, we have introduced a common physical amplitude factor for clarity:
\begin{equation}
  \mathcal{A}_{\text{phys}}(m_{A'}) = \sqrt{\frac{4\pi}{3}}\, \frac{m_{A'}^2 R}{2-(m_{A'}R)^2}.
  \label{eq:A_phys_def}
\end{equation}
The terms in the square brackets, such as $[u_\theta (\boldsymbol{\Phi}_{1,0}\cdot\hat{\theta}) + u_\phi (\boldsymbol{\Phi}_{1,0}\cdot\hat{\phi})]$, are purely geometric factors that depend on the known site location $\Omega$ and the measured background field direction $(u_\theta, u_\phi)$. We define these complex geometric factors collectively as $h_m(\Omega)$:
\begin{equation}
    h_m(\Omega) \equiv u_\theta (\boldsymbol{\Phi}_{1,m}\cdot\hat{\theta}) + u_\phi (\boldsymbol{\Phi}_{1,m}\cdot\hat{\phi}).
    \label{eq:geom_gain_h}
\end{equation}
Then the complex amplitudes can be written more compactly as
\begin{equation}
  S_{-m}(\Omega) = \mathcal{A}_{\text{phys}}\, \varepsilon\, A'_{m} \, h_{m}(\Omega),
\end{equation}
where the index on $S$ now corresponds directly to the frequency offset ($S_0 \to f_{A'}$, $S_- \to f_{A'}-f_d$, $S_+ \to f_{A'}+f_d$).

This factorization is central to our analysis. It separates the unknown physics we aim to constrain—the overall signal strength proportional to $\varepsilon$ and the stochastic polarization amplitudes $A'_m$—from the deterministic geometric factors $h_m(\Omega)$ that can be calculated from our knowledge of the experimental setup. In our likelihood framework (Sec.~\ref{sec:analysis}), the model mean will be constructed as a linear combination of these predictable geometric structures, weighted by the unknown physical amplitudes.

\section{Experimental Setup and Data Acquisition}
\label{sec:instrumentation}

Our search for dark photon DM relies on high-precision magnetic field measurements conducted in a low-noise environment. This section details the core components of our experiment: the optically pumped magnetometers, the remote experimental site, and the data acquisition system designed for synchronized, long-duration operation.

\subsection{Sensors and Site Layout}
\label{subsec:sensors_site}

The primary instruments for this work are two commercial, high-sensitivity scalar optically pumped magnetometers (OPMs), specifically the Pulsed Pump Magnetometer (PPM) model from Twinleaf LLC \cite{TwinleafPPM}. These sensors, hereafter denoted \textsf{TL-1} and \textsf{TL-2}, operate on the principle of pulsed free-precession of spin-polarized $^{87}$Rb vapor \cite{Limes2020Portable}. The measurement sequence is a three-step cycle:
\begin{enumerate}
    \item \textbf{Optical Pumping:} A short, high-power pulse of a resonant, circularly polarized pump laser polarizes the atomic spins, a technique similar to the Bell--Bloom scheme \cite{Bell1961Optical}.
    \item \textbf{Free Precession:} The pump laser is extinguished, allowing the spin ensemble to precess freely about the ambient magnetic field $\mathbf{B}_b$. The precession occurs at the Larmor frequency, $\omega_L = \gamma |\mathbf{B}_b|$, which is directly proportional to the magnitude of the magnetic field via the gyromagnetic ratio $\gamma$.
    \item \textbf{Optical Readout:} A weak, linearly polarized, and far-detuned probe laser continuously passes through the atomic vapor. The precessing spins induce an oscillation in the probe laser's polarization plane via the Faraday effect. This rotation is measured with a balanced polarimeter, and the resulting signal's frequency yields a precise measurement of the magnetic field magnitude.
\end{enumerate}
Key performance parameters of the sensors, based on manufacturer specifications and our own characterization (see Sec.~\ref{sec:systematics}), are summarized in Table~\ref{tab:sensor_params}.

To minimize anthropogenic electromagnetic interference, which is a dominant noise source for sensitive magnetometer experiments, the measurement campaign was conducted at a remote field station in the Austrian Alps. The site, located at geographic coordinates \mbox{N 47\si{\degree}9'42''}, \mbox{E 13\si{\degree}55'43''}, was specifically selected for its distance from power lines, roads, and other sources of human-made electromagnetic interference. Situated several kilometers from the nearest urban infrastructure, and with the natural topography providing additional shielding, this pristine environment is crucial for achieving the low noise floor required to search for faint, narrow-band signals predicted from dark matter.

The two magnetometer heads, \textsf{TL-1} and \textsf{TL-2}, were deployed with a baseline separation of $L \approx \SI{15}{\meter}$. This separation is large enough to mitigate local, near-field instrumental cross-talk while being small enough to ensure that both sensors experience a highly correlated far-field signal from both geophysical and potential DM sources.

\begin{figure}[t]
  \centering
  \includegraphics[width=\linewidth]{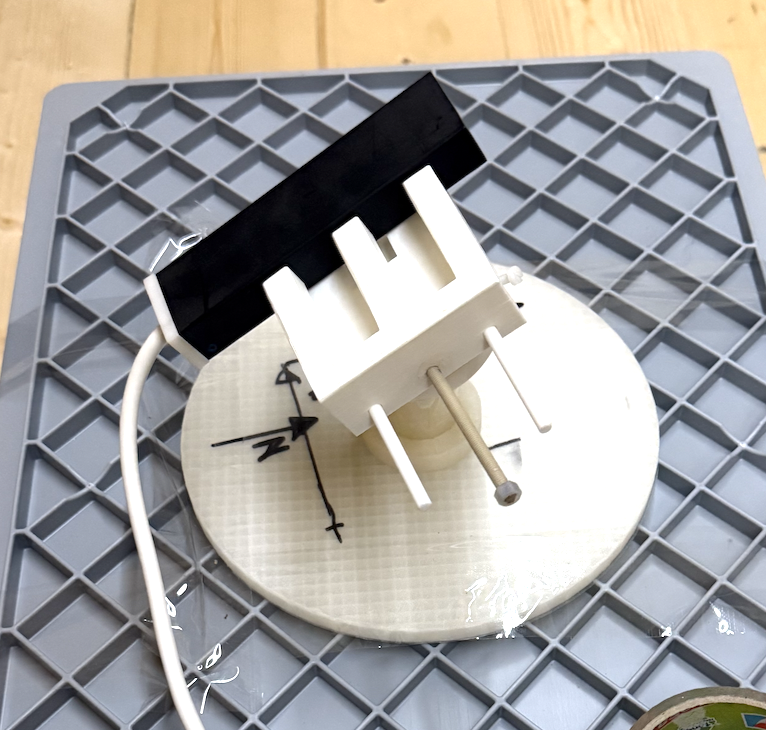}
  \caption{
  Photograph of a single Twinleaf PPM sensor head (the black rectangular unit) placed in its custom-designed, non-magnetic alignment mount. The mount, constructed from 3D-printed material, allows for precise adjustment of the sensor's orientation. The hand-drawn markings on the base serve as a reference for aligning the sensor's primary axis with the direction of the local geomagnetic field, $\hat{\mathbf{B}}_b$, a critical step for minimizing heading errors.
  }
  \label{fig:site_layout}
\end{figure}

To minimize heading errors—a systematic effect in scalar magnetometers dependent on the angle between the sensor axis and the magnetic field—each sensor was placed in a custom-designed, non-magnetic alignment mount, as shown in Fig.~\ref{fig:site_layout}. This mount allows for the precise orientation of the sensor. A key step in our procedure was to align the sensor's primary axis with the direction of the local geomagnetic field, $\hat{\mathbf{B}}_b$, as predicted by the International Geomagnetic Reference Field (IGRF) model (see Sec.~\ref{subsec:background_field_char} for details). The uncertainty in this model-based alignment procedure is a primary contributor to our systematic error budget and is accounted for in the analysis (Sec.~\ref{sec:systematics}).

\begin{table*}[t]
\caption{\label{tab:sensor_params}
Key parameters of the Twinleaf PPM sensors used in this work. The specified parameters are identical for both sensors used (\textsf{TL-1} and \textsf{TL-2}). Values marked ``meas.'' are obtained from our laboratory or in-situ characterization; others are nominal. Uncertainties represent one standard deviation. The characterization of the timebase stability is detailed in Sec.~\ref{sec:systematics}.}
\begin{ruledtabular}
\begin{tabular}{lc}
Parameter & Value \\
\hline
Nominal Sampling Rate $f_s$ [\si{\hertz}] & 480 \\
Bandwidth (3 dB) [\si{\hertz}] & $\sim 200$ \\
Noise Floor at \SI{1}{\hertz} [\si{\pico\tesla\per\sqrt{\hertz}}] & $< 0.7$ \textit{(meas.)} \\
Dynamic Range [\si{\micro\tesla}] & \SIrange{1}{100}{} \textit{(nominal)} \\
Slew Rate [\si{\nano\tesla\per\milli\second}] & $> 1000$ \textit{(nominal)} \\
Dead-Zone Half-Angle [\si{\degree}] & $< 7$ \textit{(meas.)} \\
Timebase Stability (Drift) [\si{\nano\hertz\per\second}] & $\sim 73$ \textit{(meas.)} \\
\end{tabular}
\end{ruledtabular}
\end{table*}

\subsection{Data Acquisition and Synchronization}
\label{subsec:daq_sync}

Data from both magnetometers are acquired continuously via separate USB connections to a single host laptop running a custom Python-based data acquisition interface. Power for the entire setup, including the sensors and the laptop, is supplied by a large-capacity battery bank, which is isolated from the mains and trickle-charged by solar panels during the day to ensure uninterrupted, low-noise operation.

A crucial challenge for a multi-sensor array is achieving precise time synchronization. Since the two sensors operate as independent USB peripherals, their data streams are timestamped by the host operating system upon data packet arrival. This provides a common time reference, but it is not perfectly uniform due to operating system latencies and instrumental artifacts.

A significant instrumental artifact, identified during the characterization phase of this project, is the presence of sporadic, unflagged "time jumps" in the data stream from the OPMs. These events correspond to one or more missed samples, which break the assumption of a perfectly uniform time series and can introduce critical phase errors in a Fourier analysis if not corrected. The detection and mitigation of these time jumps are therefore a key component of our data preprocessing pipeline, as detailed in Sec.~\ref{sec:data_preproc}. Successfully accounting for these artifacts is essential for coherently combining data from both sensors and for preserving the signal's integrity across long integration times.

\subsection{Background Field Characterization}
\label{subsec:background_field_char}

The scalar projection signal model, $s(t) = \hat{\mathbf{B}}_b \cdot \mathbf{B}_a(t)$, requires precise knowledge of the local geomagnetic background field vector $\mathbf{B}_b$. Specifically, our analysis depends critically on its \textbf{direction}, $\hat{\mathbf{B}}_b$, which determines the geometric gain for the projected signal. The magnitude of the background field, $|\mathbf{B}_b|$, manifests as a large, quasi-static DC offset in the raw data. As detailed in Sec.~\ref{sec:data_preproc}, this offset is removed during the data preprocessing stage and therefore does not enter the final frequency-domain analysis. Our characterization efforts thus focus on accurately determining the field's direction.

Unlike the field's magnitude, its direction cannot be directly measured by our scalar OPMs. Therefore, we rely on a well-established geophysical model to estimate the direction vector $\hat{\mathbf{B}}_b$ for our experimental location and epoch.

Specifically, we use the International Geomagnetic Reference Field (IGRF) model, which provides high-fidelity predictions of Earth's magnetic field components globally \cite{IGRF13_Alken2021}. Using the geographic coordinates of our experimental site (Sec.~\ref{subsec:sensors_site}) and its altitude, the IGRF model yields the expected declination (angle with respect to geographic north) and inclination (dip angle with respect to the horizontal). These two angles uniquely define the unit vector $\hat{\mathbf{B}}_b$ in the local geographic frame. We then perform the necessary coordinate transformation to obtain the direction cosines $(u_r, u_\theta, u_\phi)$ in the Earth-fixed spherical coordinate system required for constructing the signal model in Eq.~\eqref{eq:s_projection_tangential}. The radial component $u_r$ is also determined but does not enter the final signal projection due to the tangential nature of the dark photon field.

The reliance on a model instead of a direct in-situ measurement introduces a systematic uncertainty into our analysis, as the accuracy of the IGRF model at any specific location can be affected by local crustal anomalies and unmodeled secular variations. The uncertainty associated with this model-derived direction is a key input to our systematic error budget, as discussed in Sec.~\ref{sec:systematics}.

\section{Data Processing and Analysis Vector Construction}
\label{sec:data_preproc}

The raw data streams from the magnetometers must undergo a multi-stage processing pipeline to transform them into a format suitable for a high-sensitivity, frequency-domain search. This section begins by characterizing the raw data in both the time and frequency domains, then details the steps taken to clean and prepare the data for the construction of our final analysis vectors.

\begin{figure}[t!]
  \centering
  \includegraphics[width=\linewidth]{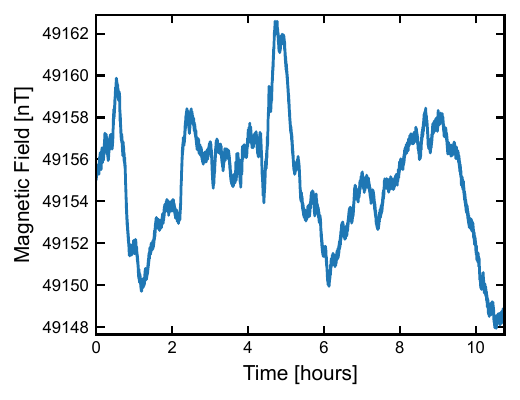}
  \caption{The raw time-domain data from one of the optically pumped magnetometer sensors, showing the full \SI{10.5}{\hour} measurement period used for this analysis. The signal is dominated by the large, quasi-static geomagnetic field (approximately \SI{49150}{\nano\tesla}). A clear diurnal variation with an amplitude of several \si{\nano\tesla} is visible, upon which high-frequency instrumental and environmental noise is superimposed. The expected dark photon signal is many orders of magnitude smaller and is therefore completely buried within the noise fluctuations.}
  \label{fig:timeseries_example}
\end{figure}

Figure~\ref{fig:timeseries_example} displays the full \SI{10.5}{\hour} segment of the raw time-domain data from one of the sensors used in this analysis. The data is dominated by Earth's quasi-static magnetic field, with a magnitude of approximately \SI{49150}{\nano\tesla}. Superimposed on this large DC offset is a clear diurnal variation on the order of several \si{\nano\tesla}, driven by solar-wind interactions with the ionosphere and magnetosphere. The dark photon signal we are searching for is expected to be many orders of magnitude smaller and is entirely buried within the high-frequency noise fluctuations visible on the plot. This illustrates the critical need for the detrending procedures described in Sec.~\ref{subsec:preprocessing}.

\begin{figure}[t!]
  \centering
  \includegraphics[width=\linewidth]{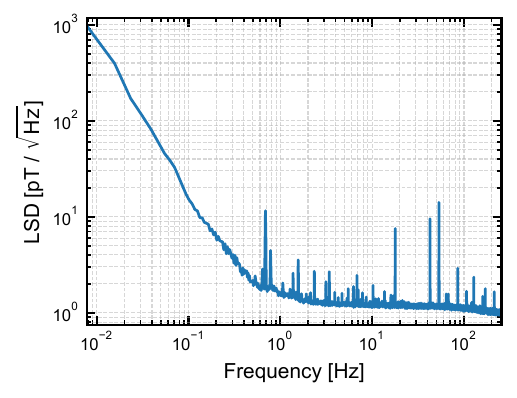}
  \caption{The Linear Spectral Density (LSD) of the data from a single representative sensor, calculated over the entire 10.5-hour observation period. The plot reveals the instrument's noise floor as a function of frequency, which consists of a broadband continuum and numerous discrete spectral lines. The continuum shows a characteristic $1/f$ behavior at low frequencies, while at higher frequencies, the narrow peaks correspond to a mixture of instrumental artifacts, which are largely unique to this specific sensor, and environmental interference, which is observed in both. This complete spectrum represents the background against which the search for a narrow dark photon signal triplet is performed.}
  \label{fig:lsd_example}
\end{figure}

Figure~\ref{fig:lsd_example} shows the frequency-domain view of our data, plotting the Linear Spectral Density (LSD) from a representative long data segment of a single sensor. This plot reveals the noise landscape of our experiment, which is discussed in detail in Appendix~\ref{app:lsd}. The spectrum consists of two main components: a broadband noise continuum and a set of discrete spectral lines. The continuum exhibits a characteristic $1/f$-like behavior at low frequencies, transitioning to a flatter, white-noise-dominated floor at higher frequencies. Superimposed on this are numerous sharp peaks, which are a mixture of stable instrumental artifacts and intermittent environmental interference. This complete noise spectrum constitutes the background against which our search for a narrow, coherent triplet signal is conducted. The goal of our analysis is to statistically determine if such a signal exists above this complex, measured noise floor.

\subsection{Preprocessing Pipeline}
\label{subsec:preprocessing}

\paragraph{Timestamp Correction and Gridding.}
\label{par:timestamp_gridding}
A prerequisite for a precise Fourier analysis is a dataset sampled on a uniform time grid. The raw data streams from the OPMs, however, can exhibit the sporadic timing discontinuities described in Sec.~\ref{subsec:daq_sync}, where one or more samples are occasionally dropped by the data acquisition system. Our preprocessing pipeline therefore includes a dedicated procedure to address these irregularities and map the time series onto a uniform grid.

The process begins by establishing a nominal sampling period, $\Delta t = 1/f_s$, calculated from the median of the inter-sample time differences over a long, stable portion of the data. The entire time series is then scanned to identify any interval where the observed time step, $\Delta t_n^{\text{obs}} = t_n - t_{n-1}$, significantly deviates from the nominal period, i.e., where $\Delta t_n^{\text{obs}} > (1+\eta)\Delta t$ for a small tolerance factor $\eta$. These discontinuities are managed based on their duration:
\begin{itemize}
    \item \textbf{Single-sample gaps:} If a discontinuity corresponds to a single missing sample, its value is estimated using linear interpolation between the two adjacent points. This minimally invasive method preserves the continuity of the data for subsequent filtering stages.
    \item \textbf{Extended gaps:} Gaps corresponding to two or more consecutive missing samples are deemed too large to be reliably interpolated. The data within these intervals are consequently flagged as invalid and are excluded from the analysis.
\end{itemize}
This systematic approach maps the raw data onto a uniform time grid, a critical foundation for the subsequent Fourier transform, while carefully accounting for any interpolated or excluded data points.

\paragraph{Despiking.}
The time series can be contaminated by sharp, impulsive spikes originating from instrumental glitches or environmental disturbances. To remove these outliers in a robust manner, we employ an iterative algorithm based on the Median Absolute Deviation (MAD). For each pass of the algorithm, we first compute a robust local baseline using a median filter and identify outliers as points that deviate from this baseline by more than a set threshold based on the local MAD. Samples flagged as outliers are then replaced using linear interpolation from their nearest valid neighbors. This interpolation method is chosen over simpler replacement strategies (such as substitution with the preceding sample) as it introduces minimal distortion to the frequency spectrum of the data. The entire process is repeated for two passes to effectively remove both primary spikes and any secondary outliers that may become apparent after the initial cleaning.

\paragraph{Detrending.}
The raw time series is dominated by large, low-frequency components that must be removed before the frequency-domain analysis. These include the quasi-static DC offset from the geomagnetic field, $|\mathbf{B}_b|$, and slow drifts caused by environmental or instrumental effects. If left uncorrected, these components would overwhelm the spectrum and obscure any potential high-frequency signal. Our detrending procedure consists of two steps: first, we subtract the mean value of the full dataset of duration $T$ to remove the DC offset. Subsequently, we fit and subtract a linear function of time, $a+bt$, to remove any residual linear drift. This process effectively acts as a high-pass filter with a cutoff frequency of $\sim 1/T$, which is well below our frequency band of interest and does not attenuate the target signal.

\subsection{Frequency-Domain Transform and the Analysis Vector}
\label{subsec:analysis_vector_construction}

After preprocessing, the data is transformed into the frequency domain to construct the analysis vector used in our likelihood framework. A critical aspect of our methodology is the choice of the Fourier transform technique, which is designed to precisely probe the signal's characteristic triplet structure.

\paragraph{Core Method: Discrete-Time Fourier Transform (DTFT).}
Traditional spectral analysis often relies on the Fast Fourier Transform (FFT), which efficiently computes the spectrum on a discrete grid of frequencies, $f_k = k/T_{\text{obs}}$, where $T_{\text{obs}}$ is the total observation time. However, the physical frequencies of the dark photon signal triplet, $\{f_{A'}, f_{A'} \pm f_d\}$, are not, in general, aligned with this FFT grid. Evaluating the signal at the nearest FFT bin introduces a systematic error known as \textbf{grid-mismatch error} or spectral leakage, which can significantly reduce sensitivity, especially for very narrow-band signals.

To quantify this issue for our analysis, we compare two critical frequency scales: the intrinsic width of the dark photon signal, $\Delta f_{\text{DM}}$, and the frequency resolution of the FFT, $\Delta f_{\text{FFT}}$. The signal's width is determined by the dark matter velocity dispersion ($v_{\text{DM}} \sim 10^{-3}$) and is given by $\Delta f_{\text{DM}} \approx 10^{-6} f_{A'}$ \cite{FedderkeSuperMAG}. In contrast, the FFT bin width is fixed by our total observation time of $T_{\text{obs}} \approx \SI{10.5}{\hour}$, yielding $\Delta f_{\text{FFT}} = 1/T_{\text{obs}} \approx \SI{26.5}{\micro\hertz}$. Throughout our entire search band, the physical signal width is thus orders of magnitude smaller than the FFT bin spacing (e.g., at $f_{A'} = \SI{1}{\hertz}$, $\Delta f_{\text{DM}} \approx \SI{1}{\micro\hertz} \ll \Delta f_{\text{FFT}}$). This extreme narrowness means the signal behaves like a delta function relative to the FFT grid. If its true frequency falls between two FFT bins, its power is dispersed across many neighboring bins, leading to a substantial loss in peak signal-to-noise ratio. Contrary to intuition, this problem becomes more severe at lower frequencies where the signal becomes even narrower relative to the fixed FFT bin width. It is precisely to mitigate this sensitivity loss that we employ the DTFT.

To circumvent this issue, we employ the Discrete-Time Fourier Transform (DTFT), which can be evaluated at any arbitrary frequency $f$. For a given data sequence $x[n]$ of length $N$ with sampling period $\Delta t$, the DTFT is defined as:
\begin{equation}
  \tilde{s}(f) = \Delta t \sum_{n=0}^{N-1} x[n]\,e^{-i2\pi f n \Delta t}.
  \label{eq:dtft_definition}
\end{equation}
This definition differs from the textbook unitless definition by the prefactor $\Delta t$. This normalization is chosen to make the DTFT a direct discrete approximation of the continuous Fourier Transform, $\int x(t) e^{-i2\pi f t} dt$. As a result, the quantity $|\tilde{s}(f)|^2 / (N\Delta t)$ gives an estimate of the Power Spectral Density (PSD) in units of $[\text{signal unit}]^2/\text{Hz}$, and $\tilde{s}(f)$ itself has units of $[\text{signal unit}]\cdot[\text{time}]$, or \si{\pico\tesla\cdot\second} in our case. This convention ensures that the amplitudes derived from the DTFT are directly comparable to theoretical predictions without requiring additional frequency-bin-width normalization factors. While direct summation is computationally expensive, we efficiently evaluate the DTFT on the required frequency arcs using the Chirp Z-Transform (CZT) algorithm \cite{Rabiner1969_CZT}, which provides the same numerical result as direct evaluation but with a computational complexity close to that of the FFT method. This approach allows us to probe the spectrum precisely at the physical frequencies of the signal model, thus eliminating grid-mismatch error and simplifying the subsequent likelihood analysis.

\paragraph{The Six-Channel Analysis Vector.}
We apply the DTFT to the preprocessed data streams from both sensors ($j=1, 2$) at the three physical frequencies of the dark photon triplet:
\begin{equation}
  f_0 \equiv f_{A'}, \quad f_{-} \equiv f_{A'} - f_d, \quad f_{+} \equiv f_{A'} + f_d.
  \label{eq:triplet_freqs_def}
\end{equation}
This procedure yields six complex numbers, which we organize into a six-dimensional complex column vector, hereafter referred to as the \textbf{analysis vector}, $X$:
\begin{equation}
  X = 
  \begin{pmatrix}
    \tilde{s}_1(f_0) \\
    \tilde{s}_1(f_{-}) \\
    \tilde{s}_1(f_{+}) \\
    \tilde{s}_2(f_0) \\
    \tilde{s}_2(f_{-}) \\
    \tilde{s}_2(f_{+})
  \end{pmatrix}.
  \label{eq:analysis_vector_Xk}
\end{equation}
This vector $X$ serves as the fundamental data object for our statistical analysis. It compactly represents all the relevant information from the dataset for the dark photon search at a given mass $m_{A'}$. The subsequent likelihood function, detailed in Sec.~\ref{sec:analysis}, is constructed based on the statistical properties of this vector, comparing the measured values to the theoretical model's predictions for the mean, $\mathbb{E}[X]$.

\subsection{Single-Segment Analysis and Coherence Time}
\label{subsec:coherence_time_assumption}

A key aspect of our analysis strategy is the treatment of the entire \SI{10.5}{\hour} dataset as a single, coherent analysis segment. This approach implies that the dark photon field's amplitude and phase are assumed to be stable over the full duration of the measurement. This simplification is justified by comparing our observation time, $T_{\text{obs}}$, with the theoretical dark matter coherence time, $\tau_c$.

The coherence time is the timescale over which the wavelike dark matter field maintains a constant phase. It is inversely proportional to the frequency spread of the signal, which is determined by the dark matter's velocity dispersion, $v_{\text{DM}} \sim 10^{-3}$, and is given by $\tau_c \sim (f_{A'} v_{\text{DM}}^2)^{-1} \approx 10^6 / f_{A'}$ \cite{FedderkeSuperMAG}. Our search focuses on a mass range corresponding to frequencies from approximately \SI{5}{\milli\hertz} to \SI{7.5}{\hertz}. For this broad band, the expected coherence time ranges from:
\begin{itemize}
    \item At the high-frequency end ($f_{A'} = \SI{7.5}{\hertz}$): $\tau_c \approx 1.3 \times 10^5\,\text{s} \approx \SI{37}{\hour}$.
    \item At the low-frequency end ($f_{A'} = \SI{5}{\milli\hertz}$): $\tau_c \approx 2 \times 10^8\,\text{s} \approx \num{6.3}\,\text{years}$.
\end{itemize}
In both extremes, and across the entire frequency range of interest, our total observation time of $T_{\text{obs}} \approx \SI{10.5}{\hour}$ is significantly shorter than the expected coherence time ($T_{\text{obs}} \ll \tau_c$). Therefore, it is a very well-founded approximation to treat the entire dataset as a single coherent block.

The physical implication of this is that we can perform the analysis with a single segment index, $k=1$. Consequently, the dark matter polarization amplitudes, $\mathbf{c}_k$, do not vary and can be treated as a single, constant (though unknown) vector, $\mathbf{c}$, throughout the measurement. This simplifies the likelihood framework considerably, as we no longer need to sum or product over multiple independent segments. The analysis effectively becomes a search for a single, coherent signal embedded in the full \SI{10.5}{\hour} data stream.

\section{Likelihood Framework}
\label{sec:analysis}

With the analysis vector $X$ constructed for the entire coherent segment, we now develop the statistical framework to test for the presence of a dark photon signal and to set upper limits on the kinetic mixing parameter $\varepsilon$. Our approach is based on a multi-channel complex Gaussian likelihood function.

\subsection{Statistical Model}
\label{subsec:statistical_model}

We model the six-component analysis vector $X$ as the sum of a deterministic signal mean vector $\mu$ and a stochastic noise vector $N$:
\begin{equation}
  X = \mu + N.
\end{equation}
Based on the Central Limit Theorem applied to the sum of many random phases in the underlying noise processes, we assume that the noise vector $N$ follows a zero-mean, circularly-symmetric complex normal distribution. The probability distribution of the data vector is given by:
\begin{equation}
\label{eq:complex_gaussian_likelihood}
\begin{aligned}
p\!\left(X \mid \mu, \Sigma\right)
&= \big(\pi^{6}\,\det \Sigma\big)^{-1}
\\[-2pt]
&\quad\times
\exp\!\left[-(X-\mu)^{\dagger}\,\Sigma^{-1}\,(X-\mu)\right].
\end{aligned}
\end{equation}
where $\mu = \mathbb{E}[X]$ is the model mean vector, and $\Sigma = \mathbb{E}[N N^\herm]$ is the $6 \times 6$ complex noise covariance matrix. The construction of $\mu$ is the subject of the next subsection, while the empirical estimation of $\Sigma$ from data is detailed in Sec.~\ref{subsec:noise_cov}.

\subsection{The Model Mean Vector \texorpdfstring{$\mu$}{mu}}
\label{subsec:model_mean_vector}

The model mean vector $\mu$ is the theoretical expectation of the analysis vector $X$ in the presence of a dark photon signal. It is derived by applying the DTFT operator to the theoretical scalar signal $s(t)$ from Eq.~\eqref{eq:s_three_lines}.

The DTFT of a single complex sinusoid $e^{-i2\pi f_i t}$ evaluated at a frequency $f_{\text{ch}}$ over a rectangular window of duration $T$ is given by $T \cdot D_N[(f_i - f_{\text{ch}})T]$, where $D_N$ is the Dirichlet kernel defined in Eq.~\eqref{eq:app_dirichlet_kernel}. Applying this to the three components of $s(t)$, the expected value of the DTFT for sensor $j$ at channel frequency $f_{\text{ch}}$ is a coherent sum of contributions from all three physical lines:
\begin{align}
\mu_j(f_{\text{ch}}) = \frac{T}{2} \Big( &S_0 D_N[(f_0-f_{\text{ch}})T] + S_- D_N[(f_--f_{\text{ch}})T] \nonumber \\
&+ S_+ D_N[(f_+-f_{\text{ch}})T] \Big).
\end{align}
The factor of $1/2$ arises from taking the real part of the complex exponential in the signal model.

This structure allows us to express the full 6D mean vector $\mu$ in a convenient factorized form that separates the unknown physics from the deterministic instrumental and geometric factors:
\begin{equation}
  \mu = \mathbb{E}[X] = \varepsilon \, M \, \mathbf{c},
  \label{eq:mu_factorized}
\end{equation}
where $\varepsilon$ is the kinetic mixing parameter, $\mathbf{c} = [A'_{0}, A'_{+1}, A'_{-1}]^\T$ is the vector of complex dark matter polarization amplitudes, and $M$ is the $6 \times 3$ \textbf{template matrix}.

The template matrix $M$ encodes all the known physics of the signal transduction and measurement process. Its columns correspond to the three polarization components ($m=0, +1, -1$), and its rows correspond to the six channels of the analysis vector (sensor 1 at $\{f_0, f_-, f_+\}$, and sensor 2 at $\{f_0, f_-, f_+\}$). Using the notation $\delta \equiv f_d T$ and the properties of the Dirichlet kernel ($D_N(0)=1$, $D_N(-x) = D_N(x)^*$), the matrix takes the explicit form:
\begin{widetext}
\begin{equation}
M = \frac{T \cdot \mathcal{A}_{\text{phys}}}{2}
\begin{pmatrix}
h_{1,0}               & h_{1,+1} D_N(\delta)^* & h_{1,-1} D_N(\delta) \\
h_{1,0} D_N(\delta)   & h_{1,+1}              & h_{1,-1} D_N(2\delta) \\
h_{1,0} D_N(\delta)^* & h_{1,+1} D_N(2\delta)^* & h_{1,-1} \\
h_{2,0}               & h_{2,+1} D_N(\delta)^* & h_{2,-1} D_N(\delta) \\
h_{2,0} D_N(\delta)   & h_{2,+1}              & h_{2,-1} D_N(2\delta) \\
h_{2,0} D_N(\delta)^* & h_{2,+1} D_N(2\delta)^* & h_{2,-1}
\end{pmatrix}.
\label{eq:template_matrix_M}
\end{equation}
\end{widetext}
Here, $\mathcal{A}_{\text{phys}}$ is the physical amplitude factor from Eq.~\eqref{eq:A_phys_def}, and $h_{j,m}$ are the complex geometric gains for sensor $j$ and polarization $m$, defined in Eq.~\eqref{eq:geom_gain_h}. This matrix is the core of our signal model in the frequency domain. It precisely quantifies how a dark photon signal with a given polarization state is expected to manifest in our six-channel measurement, including all cross-talk effects between the frequency channels induced by the finite observation time $T$.

\subsection{Noise Covariance Matrix \texorpdfstring{$\Sigma$}{Sigma}}
\label{subsec:noise_cov}

The sensitivity of our search is fundamentally determined by the noise level, which is characterized by the $6 \times 6$ complex noise covariance matrix, $\Sigma$. An accurate, data-driven estimation of this matrix is crucial for the validity of the likelihood analysis. We assume that the noise is stationary over frequency scales much larger than the triplet spacing $f_d$, allowing us to estimate the covariance at a target frequency $f_{A'}$ using data from its immediate frequency neighborhood.

\paragraph{Empirical Estimation from Sideband Frequencies.}
Our method for estimating $\Sigma$ is empirical, based on the assumption of local spectral stationarity. For each target frequency $f_{A'}$, we define a "sideband" region consisting of a set of frequency bins surrounding the central triplet $\{f_{A'}, f_{A'} \pm f_d\}$. This region is carefully chosen to be close enough to $f_{A'}$ to share its statistical properties, yet far enough to exclude any potential dark matter signal. Specifically, we define a "guard band" of a few frequency bins around the triplet which is excluded from the estimation. The full estimation window typically spans a range of tens of frequency resolution bins ($1/T$) on either side of the triplet.

From the DTFT samples computed in this sideband region, we form a set of sample noise vectors. The $(p,q)$-th element of the covariance matrix is then estimated as the sample covariance between the $p$-th and $q$-th channels of the analysis vector, averaged over the frequencies in the sideband region:
\begin{equation}
  (\Sigma)_{pq} = \left\langle X_p(f) X_q^*(f) \right\rangle_{f \in \text{sidebands}},
  \label{eq:cov_estimation}
\end{equation}
where $\langle \cdot \rangle_{f \in \text{sidebands}}$ denotes an average over the chosen set of off-signal frequencies. To enhance robustness against non-Gaussian noise spikes or unidentified narrow-band spectral lines, we use the median to estimate diagonal elements (power spectra) and a robust estimator for the off-diagonal elements (cross-spectra). This procedure is performed once across the sidebands of the full dataset, yielding a single, high-precision estimate of the noise covariance matrix $\Sigma$.

\paragraph{Structure of the Covariance Matrix.}
The structure of $\Sigma$ reflects the physical correlations between the channels. The two magnetometers, \textsf{TL-1} and \textsf{TL-2}, are physically separated by a distance $L \approx \SI{15}{\meter}$. At the frequencies of interest, instrumental noise sources are uncorrelated between the two sensors. While very long-wavelength environmental magnetic noise could in principle be correlated, our single-site setup is designed to be sensitive to a global signal, and we treat any residual correlated environmental noise as part of the overall noise budget. Consequently, the cross-terms between sensor 1 and sensor 2 in the covariance matrix are expected to be negligible.

This leads to a block-diagonal structure for $\Sigma$:
\begin{equation}
  \Sigma \approx \begin{pmatrix} \Sigma^{(1)} & \mathbf{0}_{3\times3} \\ \mathbf{0}_{3\times3} & \Sigma^{(2)} \end{pmatrix},
  \label{eq:block_diagonal_sigma}
\end{equation}
where $\Sigma^{(j)} \in \mathbb{C}^{3\times3}$ is the covariance matrix for sensor $j$ alone. The diagonal elements of $\Sigma^{(j)}$ represent the power spectral density for sensor $j$ at the three channel frequencies. The off-diagonal elements, such as $\text{Cov}(\tilde{s}_j(f_0), \tilde{s}_j(f_-))$, are generally non-zero. They capture the intrinsic correlations between the DTFT samples at different frequencies induced by the spectral leakage of the finite-time window (the Dirichlet kernel for a rectangular window). These terms are typically small but are fully captured by our empirical estimation procedure from Eq.~\eqref{eq:cov_estimation}, ensuring that our noise model is a faithful representation of the data.

\subsection{Likelihood and Parameter Inference}
\label{subsec:global_likelihood}

With the statistical model for a single coherent dataset in place, we now construct the likelihood function and detail the procedure for inferring the kinetic mixing parameter $\varepsilon$.

\paragraph{Likelihood Function.}
From the complex normal distribution in Eq.~\eqref{eq:complex_gaussian_likelihood}, the log-likelihood for the dataset is given by:
\begin{equation}
\label{eq:log_likelihood_segment}
\begin{aligned}
\ln \mathcal{L}\!\left(X \mid \varepsilon, \mathbf c\right)
&= -\,(X-\varepsilon M \mathbf c)^{\dagger}\,\Sigma^{-1}\,(X-\varepsilon M \mathbf c)
\\[-2pt]
&\quad -\,\ln\det\Sigma .
\end{aligned}
\end{equation}
where we have dropped the constant term $-\ln(\pi^6)$.

\paragraph{Marginalization over Stochastic Amplitudes.}
The complex polarization amplitude vector $\mathbf{c} = [A'_{0}, A'_{+1}, A'_{-1}]^\T$ is a stochastic quantity, treated as a single random variable for the entire observation period. In the standard halo model, it is expected to be drawn from a zero-mean, isotropic complex Gaussian distribution, reflecting the random-phase nature of the wavelike dark matter field. The normalization is set by the local dark matter density, such that $\langle \mathbf{c}^\herm \mathbf{c} \rangle = 1$.

To obtain a likelihood that depends only on the parameter of interest, $\varepsilon$, we marginalize over the unknown amplitude vector $\mathbf{c}$. This is done by integrating the likelihood over the prior probability distribution of the amplitude, $p(\mathbf{c})$. Assuming a Gaussian prior for $\mathbf{c} \sim \mathcal{CN}(0, \frac{1}{3}I_{3\times3})$, the marginalization can be performed analytically \cite{FedderkeSuperMAG, Centers2019_Stochastic}:
\begin{equation}
  \mathcal{L}(X | \varepsilon) = \int \mathcal{L}(X | \varepsilon, \mathbf{c}) p(\mathbf{c}) d^6\mathbf{c}.
\label{eq:marginalized_likelihood_k}
\end{equation}
This integral is a standard Gaussian integral, and after performing a change of variables and integrating, one arrives at a compact expression. Following the methodology of Refs.~\cite{FedderkeSuperMAG,Arza2022_Axion}, it is advantageous to first "whiten" the data and the model. Let $\Sigma = A A^\herm$ be the Cholesky decomposition of the covariance matrix. We define the whitened data $Y = A^{-1} X$ and the whitened template matrix $N = A^{-1} M$. Then, performing a singular value decomposition (SVD) on the whitened template matrix, $N = U S V^\herm$, where $S$ is a diagonal matrix of singular values $s_{i}$ ($i=1,2,3$), the marginalized log-likelihood can be expressed in terms of the projected whitened data $Z = U^\herm Y$:
\begin{equation}
  \ln \mathcal{L}(X | \varepsilon) = \sum_{i=1}^3 \left[ -\ln(3+\varepsilon^2 s_{i}^2) - \frac{3|z_{i}|^2}{3+\varepsilon^2 s_{i}^2} \right],
  \label{eq:log_marginalized_likelihood_final}
\end{equation}
plus a constant independent of $\varepsilon$.

\paragraph{Bayesian Inference and Jeffreys Prior.}
To infer $\varepsilon$, we adopt a Bayesian approach. The posterior probability density function (PDF) for $\varepsilon$ given the data $X$ is given by Bayes' theorem:
\begin{equation}
  p(\varepsilon | X) \propto \mathcal{L}(X | \varepsilon) \cdot p(\varepsilon),
\end{equation}
where $p(\varepsilon)$ is the prior PDF for $\varepsilon$. We choose a non-informative prior to reflect our lack of prior knowledge about the value of $\varepsilon$. Specifically, we use the Jeffreys prior, which is invariant under reparameterization of the parameter. For this likelihood, the Jeffreys prior is given by \cite{FedderkeSuperMAG}:
\begin{equation}
  p(\varepsilon) \propto \left( \sum_{i=1}^3 \frac{4\varepsilon^2 s_{i}^4}{(3+\varepsilon^2 s_{i}^2)^2} \right)^{1/2}.
  \label{eq:jeffreys_prior}
\end{equation}
Combining the marginalized likelihood from Eq.~\eqref{eq:log_marginalized_likelihood_final} with the Jeffreys prior from Eq.~\eqref{eq:jeffreys_prior} yields the final (unnormalized) posterior PDF for $\varepsilon$:
\begin{widetext}
\begin{equation}
  p(\varepsilon | X) \propto \left( \sum_{i=1}^3 \frac{4\varepsilon^2 s_{i}^4}{(3+\varepsilon^2 s_{i}^2)^2} \right)^{1/2} \prod_{i=1}^3 \frac{1}{3+\varepsilon^2 s_{i}^2} \exp\left(-\frac{3|z_{i}|^2}{3+\varepsilon^2 s_{i}^2}\right).
  \label{eq:posterior_pdf_final}
\end{equation}
\end{widetext}
This expression is the direct analogue of Eq.~(63) in Ref.~\cite{FedderkeSuperMAG}, adapted for our six-channel scalar magnetometer analysis. The product in our expression runs only over the singular value index $i$, whereas the expression in Ref.~\cite{FedderkeSuperMAG} includes an additional product over a segment index $k$. This simplification arises because, as detailed in Sec.~\ref{subsec:coherence_time_assumption}, our total observation time is much shorter than the dark matter coherence time, allowing us to treat the entire dataset as a single coherent analysis segment. The final step is to numerically normalize this posterior PDF and compute the credible upper limit, as detailed in Sec.~\ref{sec:results}.

\section{Results and Upper Limits}
\label{sec:results}

Having established the full Bayesian framework, we now proceed to extract the primary scientific result of this work: frequency-resolved upper limits on the kinetic mixing parameter $\varepsilon$. This is achieved by numerically evaluating the posterior PDF derived in the previous section for each target dark photon mass $m_{A'}$.

\subsection{Posterior Probability and Credible Upper Limits}
\label{subsec:posterior_and_limits}

The unnormalized posterior PDF for $\varepsilon$, given the full dataset $X$, is given by Eq.~\eqref{eq:posterior_pdf_final}. To obtain a properly normalized PDF, we first compute the normalization constant, $\mathcal{N}$, by numerically integrating the unnormalized posterior, $p_{\text{un}}(\varepsilon | X)$, over all possible values of $\varepsilon$:
\begin{equation}
  \mathcal{N} = \int_0^\infty p_{\text{un}}(\varepsilon | X) \, d\varepsilon.
\end{equation}
The integration is performed over $\varepsilon \ge 0$, as the kinetic mixing parameter is conventionally taken to be positive. The normalized posterior PDF is then simply
\begin{equation}
  p(\varepsilon | X) = \frac{1}{\mathcal{N}} p_{\text{un}}(\varepsilon | X).
\end{equation}

With the normalized posterior in hand, we can set a Bayesian credible upper limit on the parameter. We calculate the one-sided 95\% credible level (C.L.) upper limit, denoted as $\hat{\varepsilon}_{95}$, by finding the value of $\varepsilon$ that contains 95\% of the total posterior probability, starting from zero. Mathematically, $\hat{\varepsilon}_{95}$ is the solution to the integral equation:
\begin{equation}
  \int_0^{\hat{\varepsilon}_{95}} p(\varepsilon | X) \, d\varepsilon = 0.95.
  \label{eq:credible_limit_integral}
\end{equation}
This integral represents the cumulative distribution function (CDF) of the posterior, evaluated at $\hat{\varepsilon}_{95}$.

In practice, this procedure is carried out numerically for each frequency $f_{A'}$ (and corresponding mass $m_{A'}$) in our search band. For each frequency:
\begin{enumerate}
    \item We construct the required quantities $z_i$ and $s_i$ from the data and the model, as described in Sec.~\ref{sec:analysis}.
    \item We evaluate the unnormalized posterior PDF, Eq.~\eqref{eq:posterior_pdf_final}, on a fine grid of $\varepsilon$ values. The grid is chosen to be logarithmically spaced and wide enough to cover the entire region where the posterior has significant support.
    \item We compute the cumulative integral of the posterior using a numerical quadrature method (e.g., the trapezoidal rule) on the grid.
    \item Finally, we find the value $\hat{\varepsilon}_{95}$ by linearly interpolating the grid to find the point where the cumulative integral equals 0.95.
\end{enumerate}
This process is repeated for every frequency point in our scan, yielding a continuous upper limit curve, $\hat{\varepsilon}_{95}(m_{A'})$, which is the main result presented in the following subsection.

\subsection{Exclusion Limits and Comparison with Previous Work}
\label{subsec:results_and_comparison}

\begin{figure*}[t]
  \centering
  \includegraphics[width=\textwidth]{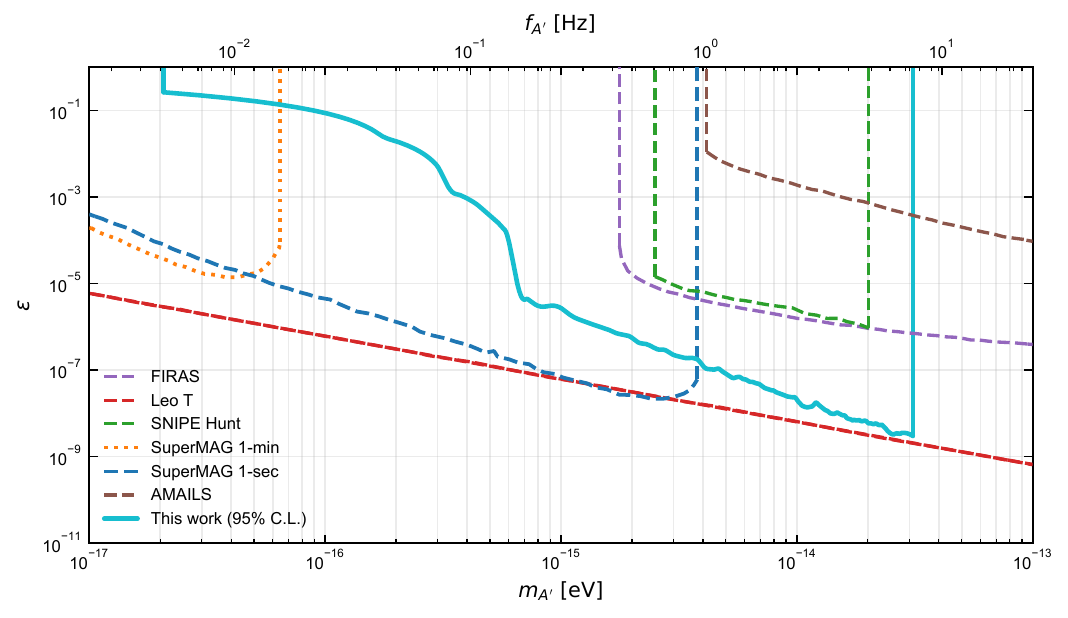}
  \caption{
  Our 95\% C.L. upper limit on the kinetic mixing parameter $\varepsilon$ as a function of the dark photon mass $m_{A'}$ (solid cyan line). 
  Excluded regions from other searches are shown for comparison, including terrestrial constraints from SNIPE Hunt \cite{SNIPE_Hunt_Placeholder}, the SuperMAG network (1-min \cite{FedderkeSuperMAG} and 1-sec \cite{FedderkeSuperMAG_1sec} datasets), and AMAILS \cite{AMAILS_Placeholder}, as well as astrophysical and cosmological constraints from FIRAS \cite{FIRAS_Placeholder} and Leo T \cite{LeoT_Placeholder}. 
  Our result establishes the most stringent direct laboratory limits to date in the mass range from approximately $4 \times 10^{-15}\,\text{eV}$ to $3 \times 10^{-14}\,\text{eV}$.
  }
  \label{fig:comparison_plot}
\end{figure*}

The primary result of our analysis is the 95\% credible level (C.L.) upper limit on the kinetic mixing parameter $\varepsilon$ as a function of the dark photon mass $m_{A'}$. This limit is presented in Fig.~\ref{fig:comparison_plot}, where our new constraint is shown as the solid cyan line, overlaid on the existing exclusion landscape for ultralight dark photons. In addition to the constraints shown in Fig.~\ref{fig:comparison_plot}, we note that a recent and independent preprint~\cite{Nomura2025_Eskdalemuir} has analyzed long-term magnetic-field data from the Eskdalemuir observatory in search of ultralight dark photon dark matter. This work provides another direct laboratory constraint based on geomagnetic observations. Besides the terrestrial and astrophysical constraints explicitly shown in Fig.~\ref{fig:comparison_plot}, many additional cosmological bounds have been derived in the literature~\cite{Caputo2020_Inhomogeneous,Bolton2022_DPDMLyman,Trost2025_DPDMLyman,McDermott2020_Cosmology}. These cosmological limits rely on specific modelling assumptions and provide information that is complementary to our local, direct laboratory constraints, and we therefore do not attempt to include them in Fig.~\ref{fig:comparison_plot}.

Our exclusion limit generally improves (i.e., $\hat{\varepsilon}_{95}$ decreases) with increasing mass at a rate faster than the simple scaling $\epsilon \propto 1/(m'_A)^2$, a trend that primarily reflects a lower intrinsic noise floor for the OPMs and a reduction in environmental magnetic noise at higher frequencies. The curve exhibits several localized features or "spikes" where the constraint becomes weaker. These correspond to frequencies with statistically significant excess power in the data that are not vetoed by our analysis. However, none of these candidates reaches a level of global significance required for a discovery claim after accounting for the look-elsewhere effect across the scanned mass range. They are consistent with statistical fluctuations or low-level, unidentified intermittent environmental noise sources.

To place these findings in the context of the global search, we compare our limit to a selection of existing constraints from other terrestrial experiments and astrophysical observations, also shown in Fig.~\ref{fig:comparison_plot}. Our result makes a significant contribution in the mass range of approximately \SIrange{4e-15}{3e-14}{\electronvolt}. In this region, our limit improves upon existing direct-detection laboratory constraints, improving upon those set by the SNIPE Hunt experiment \cite{SNIPE_Hunt_Placeholder} by up to 2.5 orders of magnitude and by an even greater factor compared to the AMAILS experiment \cite{AMAILS_Placeholder}.

Crucially, our work establishes the most stringent direct laboratory limits to date across this mass range, surpassing recent constraints from experiments such as AMAILS \cite{AMAILS_Placeholder}. This is a key contribution, as it provides the first robust, direct laboratory constraint in a parameter space that was previously covered primarily by astrophysical observations (e.g., from Leo T \cite{LeoT_Placeholder}). Because direct-detection limits are subject to a completely different and more controlled set of systematic uncertainties than their astrophysical counterparts, our result provides an important independent verification and strengthens the exclusion in this well-motivated region of the dark photon parameter space.

\section{Systematic Uncertainties and Validation}
\label{sec:systematics}

While our statistical framework is designed to be robust, several sources of systematic uncertainty can affect the final limit on $\varepsilon$. A thorough understanding and quantification of these effects are essential for the credibility of the result. In this section, we analyze the dominant systematic uncertainties and present validation tests for our analysis pipeline.

\subsection{Systematic Error Budget}
\label{subsec:systematic_budget}

\paragraph{Uncertainty from Background Field Direction.}
The core of our signal model relies on projecting the vectorial dark photon field onto the local geomagnetic field direction, $\hat{\mathbf{B}}_b$. Any uncertainty in the determination of $\hat{\mathbf{B}}_b$ directly propagates to an uncertainty in the geometric gains $h_{j,m}$ [Eq.~\eqref{eq:geom_gain_h}] and, consequently, to an uncertainty in the overall signal amplitude used in our likelihood fit. This constitutes a systematic uncertainty on our final limit $\hat{\varepsilon}$.

The direction of $\hat{\mathbf{B}}_b$ is defined by the local direction cosines $(u_\theta, u_\phi)$ in the Earth-fixed spherical basis. As described in Sec.~\ref{subsec:background_field_char}, these direction cosines are derived from the IGRF model, not from an in-situ measurement. Consequently, the uncertainty in these direction cosines arises from two main sources:
\begin{enumerate}
    \item \textbf{IGRF Model Uncertainty:} The IGRF model has inherent limitations and does not perfectly capture local crustal magnetic anomalies or short-term geomagnetic fluctuations. This contributes a baseline uncertainty to the predicted direction.
    \item \textbf{Alignment Uncertainty:} This is the physical pointing error associated with aligning the sensor's axis to the direction prescribed by the IGRF model (see Appx.~\ref{app:calibration_details} for details on the heading error characterization).
\end{enumerate}
We combine these effects into a single effective covariance matrix, $C_u$, for the direction cosines $[u_\theta, u_\phi]$.

To quantify the impact on our signal model, we perform a linear error propagation. The geometric gains $h_{j,m}$ are functions of $(u_\theta, u_\phi)$, so their first-order uncertainty is described by the covariance matrix:
\begin{equation}
  \text{Cov}(h_{j,m}, h_{j,m'}) = J_j C_u J_j^\herm,
\end{equation}
where $J_j$ is the $3 \times 2$ complex Jacobian matrix with elements $(J_j)_{m, \alpha} = \partial h_{j,m} / \partial u_\alpha$ for $m \in \{-1, 0, +1\}$ and $\alpha \in \{\theta, \phi\}$. We have derived the analytical form of this Jacobian from the definition of the vector spherical harmonics.

The most relevant quantity for our analysis is the resulting relative uncertainty on the \emph{magnitude} of the geometric gains, $\sigma(|h_{j,m}|) / |h_{j,m}|$, as this directly translates to a scaling uncertainty on the signal strength for each polarization component. We have evaluated this propagation for several scenarios representing different assumptions about the uncertainty in $(u_\theta, u_\phi)$. Assuming a conservative covariance matrix for the direction cosines, $C_u$, characterized by standard deviations of $\sigma_{u_\theta} \approx \sigma_{u_\phi} \approx \sqrt{2.5 \times 10^{-3}} \approx 0.05$ (corresponding to an angular uncertainty of a few degrees) and a strong positive correlation, our analysis yields the following typical relative uncertainties on the amplitude of the geometric gains:
\begin{itemize}
    \item For the transverse polarization components ($m=\pm 1$): The relative uncertainty is approximately \textbf{11\%}.
    \item For the longitudinal component ($m=0$): The relative uncertainty is significantly larger, at approximately \textbf{122\%}.
\end{itemize}
The much larger uncertainty for the $m=0$ component is due to the geometric suppression of this term at our site's latitude, where $|h_{j,0}|$ is intrinsically small. While this large relative uncertainty seems alarming, its overall impact on the final limit is moderated because the search sensitivity is typically dominated by the stronger transverse ($m=\pm 1$) components. We incorporate this uncertainty as a systematic error on the overall signal normalization when interpreting our final limits. A more aggressive assumption on pointing accuracy, with $\sigma_{u_\theta, u_\phi} \approx \sqrt{4.0 \times 10^{-4}} \approx 0.02$, reduces these uncertainties to 4\% and 49\%, respectively.

\paragraph{Timestamp Uncertainty and Coherence.}
A stable and accurate timebase is fundamental to any search for narrow-band signals. We identify two primary sources of timing uncertainty: long-term clock drift and discrete sample-dropping events, or "time jumps."

Long-term drift in the sensor's internal sampling clock was characterized by tracking the apparent frequency of a stable, injected 33-Hz tone over 18.5 hours. Our characterization of the timebase, performed by tracking the fitted peak frequency over time, reveals a very slow, linear drift on the order of \SI{73}{\nano\hertz\per\second}, which corresponds to a change of one Hertz over approximately 159 days. For the duration of our analysis (under 11 hours), this drift is negligible and does not contribute significantly to spectral broadening (the characterization procedure is detailed in Appx.~\ref{app:calibration_details}).

A critical instrumental artifact identified during our characterization is that the sensors are prone to sporadically dropping one or more samples without flagging an error. If uncorrected, such a phase discontinuity would spread a narrow-band signal's power across the frequency domain, severely degrading the peak sensitivity of the DTFT. Our mitigation strategy, detailed in Sec.~\ref{subsec:preprocessing}, involves algorithmically detecting these jumps and repairing them via interpolation where possible. Gaps too large to repair would compromise the coherence of the entire dataset, effectively invalidating the single-segment assumption for that portion of data. In this analysis, no such unrepairable gaps were present in the final dataset, thus preserving the integrity of our coherent analysis.

\paragraph{Amplitude Calibration Uncertainty.}
The overall normalization of our signal model, and thus the inferred limit on $\varepsilon$, is directly proportional to the calibrated gain of the magnetometers. An uncertainty in the gain translates linearly into an uncertainty on $\hat{\varepsilon}$.

We characterized the coherent gain of each sensor in a magnetically shielded environment by injecting a weak, sinusoidal magnetic field of known amplitude at a reference frequency within our search band (e.g., \SI{33}{\hertz}). The gain is determined by the ratio of the measured signal amplitude in the spectrum to the known amplitude of the injected field. The procedure was repeated at several frequencies to confirm that the gain is flat across our band of interest (see Appx.~\ref{app:calibration_details} for the full calibration methodology).

The dominant sources of uncertainty in this calibration are the accuracy of the function generator producing the driving current and the geometric factor of the calibration coil. Combining these effects, we estimate the total systematic uncertainty on the coherent gain of each sensor to be \textbf{2\%}. Since the limit $\hat{\varepsilon}$ is inversely proportional to the measured signal amplitude, this 2\% gain uncertainty propagates directly as a 2\% systematic uncertainty on our final reported exclusion limit. We have not observed any significant dependence of the sensor gain on the absolute ambient field strength in the relevant range, although the noise floor does show such a dependence (as discussed in the \hyperref[par:noise_model_validation]{paragraph on Noise Model Validation}).

\paragraph{Noise Model Validation: Local Stationarity and Gaussianity.}
\label{par:noise_model_validation}
Our likelihood framework, Eq.~\eqref{eq:complex_gaussian_likelihood}, relies on two core assumptions about the noise: that it is a complex Gaussian process, and that its statistical properties (specifically its covariance) are stationary over the local frequency scale used for its empirical estimation. We performed a series of statistical tests on the data from the frequency sidebands to validate these assumptions.

First, we tested for \textbf{local stationarity} by comparing the covariance matrices estimated from frequency bands to the left and right of the central signal triplet. Using a multivariate generalization of Levene's test known as Box's M-test, we found no statistically significant difference between the two covariance structures ($p \approx 1$), supporting the assumption that our local sideband estimation procedure for $\Sigma$ is valid.

Second, we tested the core assumption of \textbf{complex Gaussianity}. We collected noise samples from the combined sideband region and whitened the data using the empirically estimated covariance matrix $\Sigma$. If the noise is truly complex Gaussian, the components of the resulting whitened noise vector should be independent and identically distributed according to a standard complex normal distribution. We performed a D'Agostino-Pearson $K^2$ test for normality on the real and imaginary parts of each of the six whitened channels. The combined p-values, aggregated using Fisher's method, were found to be $p_{\text{real}} = 0.38$ and $p_{\text{imag}} = 0.039$. Both values are well above our chosen significance threshold of $\alpha=0.01$, providing strong statistical support for the Gaussian hypothesis.

\begin{figure*}[t]
  \centering
  \includegraphics[width=\textwidth]{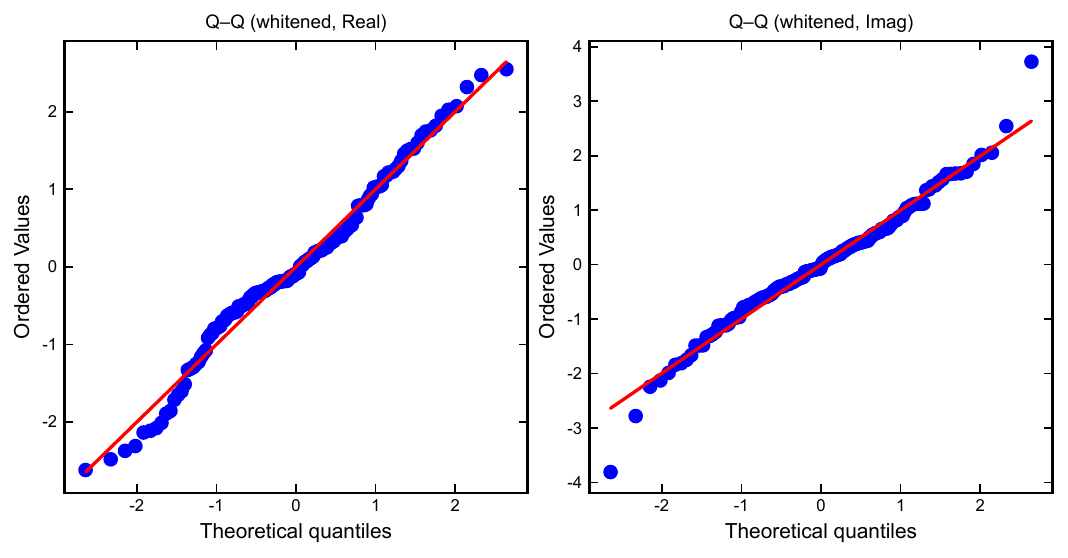}
  \caption{Quantile-Quantile (Q-Q) plots testing the Gaussianity of the whitened noise data, aggregated from the frequency sidebands around a representative central frequency. The left panel shows the real part and the right panel shows the imaginary part. The empirical quantiles of the data (blue dots) closely follow the theoretical quantiles of a standard normal distribution (red line), providing strong visual support for the complex Gaussian noise assumption central to our likelihood framework.}
  \label{fig:qq_plot}
\end{figure*}

This conclusion is further supported by a visual inspection of the Quantile-Quantile (Q-Q) plots, shown in Fig.~\ref{fig:qq_plot}. The plots for both the real and imaginary parts of the aggregated whitened data show a strong linear alignment between the empirical quantiles and the theoretical quantiles of a standard normal distribution. In summary, these validation tests confirm that our noise model provides a faithful representation of the data's statistical properties, thereby ensuring the robustness and reliability of our final derived limits.

\subsection{Validation with Signal Injection}
\label{subsec:validation_injection}

To validate the integrity and sensitivity of our entire analysis pipeline, we performed a signal injection test. This procedure involves adding a simulated dark photon signal of known parameters to the real, raw magnetometer data and then running the full analysis chain to verify that the injected signal is recovered with the expected statistical significance. This serves as an end-to-end verification of our methodology, from data preprocessing to the final statistical inference.

\paragraph{Injection Procedure.}
We generated a synthetic dark photon signal time series, $s_{\text{inj}}(t)$, following the theoretical model described in Sec.~\ref{sec:theory}. The signal was created with a specific dark photon mass, $m_{A',\text{inj}}$, corresponding to a frequency of $f_{A',\text{inj}} = \SI{1.624523}{\hertz}$. The kinetic mixing parameter for the injection, $\varepsilon_{\text{inj}}$, was set to be three times the expected 95\% C.L. upper limit at this frequency, ensuring a signal with a high signal-to-noise ratio that should be clearly recovered by our pipeline. The polarization amplitude vector $\mathbf{c}$ was simulated as a single random draw from a Gaussian distribution, consistent with the standard halo model, and held constant for the duration of the signal. This synthetic time series was then added sample-by-sample to the raw, unprocessed data streams from both magnetometers before any filtering or detrending was applied.

\paragraph{Recovery and Statistical Significance.}
The combined dataset (real noise + injected signal) was processed through the exact same analysis pipeline used for the main search. This includes all steps of preprocessing, DTFT evaluation, and the likelihood analysis. To search for candidate signals, we employ a generalized likelihood ratio test (GLRT) \cite{Kay1998_DetectionTheory}, which computes a test statistic (TS) for the signal hypothesis ($\varepsilon > 0$) versus the null hypothesis ($\varepsilon = 0$) at each frequency. The TS is defined as $\mathrm{TS} = 2(\ln\mathcal{L}_{\text{max}} - \ln\mathcal{L}_0)$, where $\mathcal{L}_{\text{max}}$ is the likelihood maximized over $\varepsilon$ and $\mathcal{L}_0$ is the likelihood evaluated at $\varepsilon=0$. For a single trial, the TS follows a chi-squared distribution, allowing for the conversion to a local statistical significance.

\begin{figure*}[t]
  \centering
  \includegraphics[width=\textwidth]{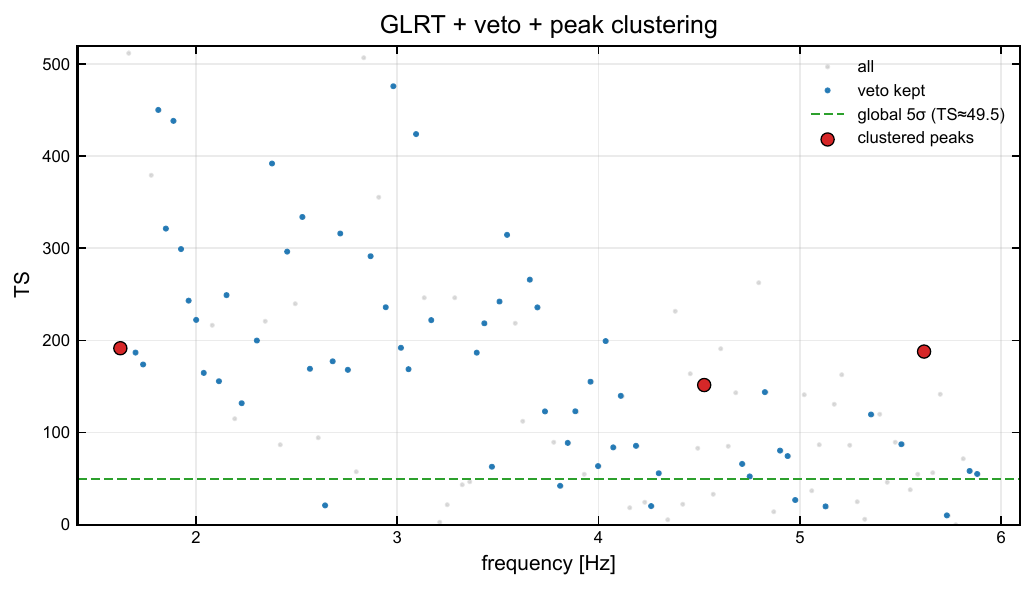}
  \caption{
  Result of the signal injection test, demonstrating the end-to-end validation of our analysis pipeline. The plot shows the test statistic (TS) for all initial candidates (light gray points). Candidates that pass our veto criteria are shown in blue, and the final, most significant peaks identified by a clustering algorithm are highlighted as large red circles. The horizontal dashed green line represents the global $5\sigma$ discovery threshold (TS $\approx 49.5$). A simulated dark photon signal, injected at $f_{A',\text{inj}} \approx \SI{1.6}{\hertz}$, is successfully recovered as a highly significant candidate (the leftmost red circle), with a TS value far exceeding the threshold.
  }
  \label{fig:injection_recovery}
\end{figure*}

Figure~\ref{fig:injection_recovery} shows the results of the GLRT scan on the injected dataset. The plot displays the TS value for each frequency in the search band. A global significance threshold, corresponding to $5\sigma$ after accounting for the look-elsewhere effect, is indicated by the dashed green line. As is evident, a prominent peak with a TS value far exceeding the discovery threshold is clearly visible at a frequency consistent with the injected signal frequency, $f_{A',\text{inj}}$.

\paragraph{Candidate Selection and Vetoing.}
To automate the identification of significant peaks and reject spurious signals (including instrumental artifacts), we employ a multi-stage candidate selection algorithm, the logic of which is implemented in our analysis code. The procedure begins by applying a series of consistency vetoes to every frequency bin. These checks are designed to ensure that any excess power conforms to the expected physical properties of a dark photon signal, including: (i) a high signal projection efficiency ($\eta$) onto the template subspace; (ii) a good quality of fit, assessed via the p-value of the whitened residual; (iii) a consistent power ratio between the central peak and sidebands of the triplet; and (iv) a physically plausible signal phase. Candidates that fail a minimum number of these checks are discarded from further consideration.

The surviving candidates, which are shown as blue points in Fig.~\ref{fig:injection_recovery}, are then passed to a peak identification and clustering pipeline. First, a global significance threshold is applied, corresponding to a $5\sigma$ discovery claim after accounting for the look-elsewhere effect (the trial factor), represented by the horizontal dashed line in the figure. Only candidates that are also local maxima (i.e., having a TS value greater than their immediate neighbors) above this threshold are considered further. To reject candidates in regions of high, fluctuating noise, we impose a local contrast requirement, ensuring a peak's TS is significantly higher than the median TS of its surrounding frequency neighborhood. Finally, a clustering algorithm groups any significant points that are within a few frequency bins of each other, selecting only the single point with the highest TS from each cluster to represent a distinct candidate peak. These final candidates are highlighted as large red circles. As shown in Fig.~\ref{fig:injection_recovery}, our injected signal successfully passes all veto and selection criteria, and is correctly identified as the most significant candidate in the entire search band. The recovered frequency is within two frequency bins of the true injected frequency, a deviation consistent with the spectral resolution of our analysis.

The candidate at approximately \SI{5.6}{\hertz}, which has a only slightly smaller TS value, is also correctly identified as a statistically significant peak by our pipeline. In a blind search, both candidates would be flagged for further investigation. However, the primary goal of this signal injection test is to validate the end-to-end recovery of a signal with known parameters. To this end, our candidate selection algorithm is configured to explicitly prioritize any significant candidate found within a narrow frequency window around the known injection frequency of $f_{A',\text{inj}}$. The successful identification of the candidate at \SI{1.6}{\hertz} under this procedure serves as the key validation of our analysis pipeline's sensitivity and correctness.

\section{Conclusion}
\label{sec:conclusion}

In this work, we have presented a comprehensive search for ultralight dark photon dark matter using a novel experimental and analytical approach. We deployed a minimalist yet powerful array consisting of two high-sensitivity scalar optically pumped magnetometers at a single, electromagnetically quiet site. By treating the wavelike dark matter signal as a projection onto the local geomagnetic field, we developed a robust six-channel likelihood framework based on the discrete-time Fourier transform, which directly probes the characteristic frequency triplet of the dark photon signal while avoiding common spectral analysis pitfalls.

Our analysis of the collected data reveals no statistically significant evidence for a dark photon signal. Consequently, we have set new world-laboratory-leading limits on the kinetic mixing parameter $\varepsilon$ in the dark photon mass range of approximately \SIrange{4e-15}{3e-14}{\electronvolt}. Our main result is a new laboratory-based exclusion limit that improves upon previous direct-detection constraints from the SNIPE Hunt experiment \cite{SNIPE_Hunt_Placeholder} by up to 2.5 orders of magnitude, and by an even greater factor compared to the AMAILS experiment \cite{AMAILS_Placeholder}.

The success of this search demonstrates the powerful potential of deploying even a small number of state-of-the-art scalar magnetometers in a well-controlled environment, when combined with a sophisticated, signal-specific analysis framework. It is particularly noteworthy that with only 10.5 hours of data, we have established limits that are competitive with those from global networks utilizing vastly larger datasets, highlighting the remarkable efficiency of this minimalist but highly sensitive experimental approach.

Looking forward, several avenues for improvement are evident. Longer-duration measurements will naturally enhance the sensitivity, particularly at lower frequencies. The inclusion of additional magnetometers could allow for more sophisticated noise cancellation techniques and provide further cross-checks on potential signal candidates. Extending this concept to a network of single-site scalar magnetometer stations, synchronized via GPS, could enable searches for large-scale correlations, combining the sensitivity of dedicated experiments with the powerful veto capabilities of a global network. Finally, further refinements to the noise modeling and a deeper investigation of instrumental systematics will continue to push the frontiers of what is achievable with this minimalist but highly effective experimental approach.

\begin{acknowledgments}
We acknowledge support by the 
Deutsche Forschungsgemeinschaft (DFG, German Research Foundation) under
Germany’s Excellence Strategy – EXC-2094 – 390783311. 
The work of YVS was supported in part by the Australian Research Council under the Discovery Early Career Researcher Award No. DE210101593.
\end{acknowledgments}

\appendix

\section{DTFT, CZT, and the Dirichlet Kernel}
\label{app:dft_details}

This appendix provides the mathematical details of the frequency-domain tools used in our analysis, as introduced in Sec.~\ref{sec:data_preproc}. We define the Discrete-Time Fourier Transform (DTFT) with our chosen normalization, derive the Dirichlet kernel as the frequency response of the implicit rectangular window, and briefly introduce the Chirp Z-Transform (CZT) as its efficient implementation.

\subsection{The Discrete-Time Fourier Transform (DTFT)}
\label{app:dtft_def}

The DTFT is the fundamental transform that maps a discrete-time sequence, $x[n]$, to a continuous function of frequency, $f$. For a finite-length sequence of $N$ samples, taken with a sampling period $\Delta t = 1/f_s$, we define the DTFT as:
\begin{equation}
  \tilde{s}(f) = \Delta t \sum_{n=0}^{N-1} x[n]\,e^{-i2\pi f n \Delta t}.
  \label{eq:app_dtft_definition}
\end{equation}
This definition differs from the textbook unitless definition by the prefactor $\Delta t$. This normalization is chosen to make the DTFT a direct discrete approximation of the continuous Fourier Transform, $\int x(t) e^{-i2\pi f t} dt$. As a result, the quantity $|\tilde{s}(f)|^2 / (N\Delta t)$ gives an estimate of the Power Spectral Density (PSD) in units of $[\text{signal unit}]^2/\text{Hz}$, and $\tilde{s}(f)$ itself has units of $[\text{signal unit}]\cdot[\text{time}]$, or \si{\pico\tesla\cdot\second} in our case. This convention ensures that the amplitudes derived from the DTFT are directly comparable to theoretical predictions without requiring additional frequency-bin-width normalization factors.

\subsection{The Dirichlet Kernel: Frequency Response of a Rectangular Window}
\label{app:dirichlet}

Our analysis is performed on finite-length data segments, which is mathematically equivalent to multiplying an infinite data stream by a rectangular window function, $w[n]=1$ for $n=0, \dots, N-1$, and zero otherwise. The frequency-domain representation of this windowing operation is a convolution of the true signal spectrum with the Fourier transform of the window function. The DTFT of the rectangular window is known as the Dirichlet kernel.

Applying our DTFT definition from Eq.~\eqref{eq:app_dtft_definition} to the window function $w[n]=1$ (and normalizing by $N$ for a unit DC response) gives:
\begin{equation}
  D_N(f) = \frac{\Delta t}{N\Delta t} \sum_{n=0}^{N-1} e^{-i2\pi f n \Delta t} = \frac{1}{N} \sum_{n=0}^{N-1} \left(e^{-i2\pi f \Delta t}\right)^n.
\end{equation}
This is a geometric series which sums to:
\begin{equation}
  D_N(f) = \frac{1}{N} \frac{1 - e^{-i2\pi f N \Delta t}}{1 - e^{-i2\pi f \Delta t}}.
\end{equation}
By factoring out phase terms from the numerator and denominator, this can be written in its more common, symmetric form:
\begin{equation}
  D_N(f) = e^{-i\pi f (N-1) \Delta t} \frac{\sin(\pi f N \Delta t)}{N \sin(\pi f \Delta t)}.
  \label{eq:app_dirichlet_kernel}
\end{equation}
In the main text, we express this kernel as a function of the normalized frequency offset $\Delta = (f_i - f_{\text{ch}})T$, where $T=N\Delta t$ is the segment duration. The phase term $e^{-i\pi f (N-1) \Delta t}$ represents the linear phase shift associated with the fact that the rectangular window is centered at $(N-1)/2$ in time, not at the origin. The magnitude $|D_N(f)|$ has a main lobe centered at $f=0$ with a width of $\sim 1/T$, and a series of decaying sidelobes. It is precisely these sidelobes that cause spectral leakage, allowing a signal at one frequency $f_i$ to contribute power when the spectrum is evaluated at a different frequency $f_{\text{ch}}$. This effect is fully captured in our template matrix $M_k$ [Eq.~\eqref{eq:template_matrix_M}] through the evaluation of the Dirichlet kernel at frequency offsets of $\pm f_d$ and $\pm 2f_d$.

\subsection{The Chirp Z-Transform (CZT)}
\label{app:czt}

Directly calculating the DTFT sum in Eq.~\eqref{eq:app_dtft_definition} for $M$ arbitrary frequency points and an $N$-point data segment has a computational complexity of $\mathcal{O}(MN)$. For the large datasets and fine frequency scans used in this work, this would be prohibitively slow. The Fast Fourier Transform (FFT) offers a much faster $\mathcal{O}(N\log N)$ complexity, but it is restricted to evaluating the spectrum on a fixed, uniform grid of frequencies $f_k = k/T$.

The Chirp Z-Transform (CZT) is an efficient algorithm that bridges this gap \cite{Rabiner1969_CZT}. It can compute the Z-transform (of which the DTFT is a special case) along a general spiral or arc-shaped contour in the complex z-plane. For our purposes, this means it can evaluate the DTFT on any set of \emph{equally spaced} frequencies, even if they are not on the FFT grid. Our three frequency channels $\{f_{A'}, f_{A'} \pm f_d\}$ for a given $f_{A'}$ are not equally spaced, but when we scan over a band of $f_{A'}$ values, we choose a fine, uniform grid for $f_{A'}$. The CZT can then be used to compute the DTFT for all three frequency-triplet-tracks simultaneously and efficiently.

The algorithm works by converting the DTFT sum into a convolution using Bluestein's algorithm, which relies on the identity $2nk = n^2 + k^2 - (k-n)^2$. This convolution can then be computed rapidly using FFTs. The resulting computational complexity is on the order of $\mathcal{O}((N+M)\log(N+M))$, which is nearly as fast as the FFT while offering the crucial flexibility of evaluating the spectrum at arbitrary starting frequencies and spacings. The use of the CZT is therefore what makes our "no grid-mismatch" analysis approach computationally feasible for searches involving large datasets.

\section{Linear Spectral Density for Noise Characterization}
\label{app:lsd}

The Linear Spectral Density (LSD) is a standard and highly effective metric for characterizing the frequency-dependent noise floor of time-series instruments, such as the magnetometers used in this work. It is derived from the more fundamental Power Spectral Density (PSD), which quantifies the distribution of a signal's power into its frequency components.

The one-sided PSD, $S_{xx}(f)$, for a real-valued time series $x(t)$ of duration $T$ is typically estimated from its discrete-time Fourier transform $\tilde{s}(f)$ (defined in Appendix~\ref{app:dft_details}) as:
\begin{equation}
  S_{xx}(f) = \frac{2}{T} \left| \tilde{s}(f) \right|^2, \quad \text{for } f > 0.
\end{equation}
The units of the PSD are signal power per unit frequency, which in our case is \si{\pico\tesla\squared\per\hertz}. While the PSD is a rigorous statistical measure, its quadratic units can be unintuitive for direct comparison with a signal's amplitude.

The LSD, $L_x(f)$, resolves this by being defined as the square root of the PSD:
\begin{equation}
  L_x(f) \equiv \sqrt{S_{xx}(f)}.
\end{equation}
This simple transformation yields two significant advantages, making the LSD an ideal tool for sensor performance analysis and comparison:

\begin{enumerate}
    \item \textbf{Intuitive Physical Units:} The units of the LSD are the signal's physical units per square-root-Hertz, in our case \si{\pico\tesla\per\sqrt{\hertz}}. The numerator (\si{\pico\tesla}) directly matches the units of the magnetic field signal we are searching for. This allows for a more direct and intuitive interpretation of the noise floor as an "amplitude spectral density." One can think of the LSD at a given frequency $f$ as the typical noise amplitude present within a \SI{1}{\hertz} bandwidth centered at $f$.

    \item \textbf{Direct Performance Comparison:} The LSD provides a standardized basis for comparing the sensitivity of different sensors. A sensor with a lower LSD value at a particular frequency has a lower intrinsic noise floor and is therefore more sensitive to faint signals at that frequency. The plot of the LSD versus frequency, as shown in Fig.~\ref{fig:lsd_example}, serves as a comprehensive "fingerprint" of the instrument's performance in the operational environment.
\end{enumerate}

In summary, by presenting the noise characteristics in terms of the LSD, we provide a clear, standardized, and physically intuitive measure of the sensitivity achieved in our experiment.

\section{Empirical Estimation of the Covariance Matrix}
\label{app:covariance_estimation}

This appendix details the practical implementation of the empirical estimation of the noise covariance matrix $\Sigma_k$, a critical input for the likelihood function in Eq.~\eqref{eq:complex_gaussian_likelihood}. The method relies on the assumption of local spectral stationarity, which posits that the statistical properties of the noise are constant over frequency scales much larger than the signal triplet's spacing. This allows us to use data from the frequency neighborhood of our target signal to characterize the noise at the signal's precise location in frequency space.

\subsection{Defining the Sideband Estimation Region}
\label{app:sideband_def}

For each dark photon mass $m_{A'}$ under consideration, corresponding to a central frequency $f_{A'}$, our goal is to estimate the $6 \times 6$ noise covariance matrix. To do this, we define a "sideband" region in the frequency domain that is close to, but does not include, the signal triplet itself. The procedure is as follows:

\begin{enumerate}
    \item \textbf{Signal Region:} First, we identify the central signal region occupied by the frequency triplet $\{f_{A'}, f_{A'} \pm f_d\}$.
    
    \item \textbf{Guard Band:} To prevent any potential, albeit weak, signal power from leaking into our noise estimate, we define a "guard band" around each of the three triplet frequencies. This typically consists of excluding a few frequency resolution bins on either side of each line. This ensures that the main lobes and nearest strong sidelobes of any potential signal are not included in the noise sample.
    
    \item \textbf{Estimation Window:} The noise samples are drawn from a wider "estimation window" that surrounds the guarded signal region. The width of this window, typically spanning tens of frequency resolution bins (e.g., $\pm 40$ bins), is chosen to be large enough to gather a sufficient number of statistically independent samples, but small enough that the assumption of noise stationarity holds.
\end{enumerate}

The final set of frequencies, $\mathcal{F}_{\text{sideband}}$, used for the estimation is therefore the set of all frequency bins within the estimation window, excluding those that fall inside the guard band.

\subsection{Robust Estimation Procedure}
\label{app:robust_estimation}

Real-world data is often contaminated by non-Gaussian noise spikes or unforeseen narrow-band interference from instrumental or environmental sources. A simple sample covariance can be heavily biased by such outliers. To ensure a robust estimate of the covariance, we employ a multi-stage procedure.

\paragraph{Outlier Rejection.}
For a given frequency $f_{A'}$, we first collect the set of $K$ analysis vectors $\{X(f_i)\}$ for all frequencies $f_i \in \mathcal{F}_{\text{sideband}}$. We then perform an outlier rejection step. For each 6D complex sample vector $X(f_i)$, we compute its Euclidean norm, $\|X(f_i)\|$. We then calculate the median and the MAD of these norms. Any sample vector whose norm deviates from the median by more than a set threshold (e.g., $3.5$ times the MAD-based robust standard deviation) is flagged as an outlier and removed from the sample set. The choice of this numerical threshold is an empirical one, set to effectively identify anomalous spectral features that could bias the noise estimate. This frequency-domain outlier rejection is a distinct processing stage from the time-domain despiking described in Sec.~\ref{subsec:preprocessing}; their respective thresholds are chosen independently to best suit their specific targets (spectral lines versus impulsive time-domain spikes). This step effectively eliminates strong, localized spectral contaminants from the noise estimation.

\paragraph{Sample Covariance Calculation.}
After removing the identified outliers, we are left with a set of $K' \le K$ clean noise vectors. The covariance matrix $\hat{\Sigma}$ is then computed as the standard sample covariance of this cleaned set:
\begin{equation}
  (\hat{\Sigma})_{pq} = \frac{1}{K'-1} \sum_{i=1}^{K'} \left( X_p(f_i) - \bar{X}_p \right) \left( X_q(f_i) - \bar{X}_q \right)^*,
  \label{eq:app_sample_covariance}
\end{equation}
where $X_p(f_i)$ is the $p$-th component of the analysis vector at frequency $f_i$, and $\bar{X}_p$ is the sample mean of the $p$-th component over the $K'$ clean samples.

This data-driven procedure has several key advantages. It is inherently adaptive, learning the noise properties directly from the data adjacent to each search frequency. It is robust against contamination by strong, non-Gaussian artifacts. Most importantly, this empirical approach automatically captures all sources of noise and correlation present in the data, including not only the intrinsic noise of the sensors and the environmental background, but also the deterministic cross-channel correlations induced by the spectral leakage of the finite-time window (i.e., the off-diagonal terms in the Dirichlet kernel response). This ensures that the covariance matrix $\Sigma_k$ used in our likelihood is a high-fidelity representation of the actual noise properties of our measurement.

\section{Instrumentation Calibration Details}
\label{app:calibration_details}

The accuracy of our final limit on $\varepsilon$ relies on a precise understanding of the magnetometers' response to both the signal and instrumental artifacts. This appendix provides further details on the key calibration procedures performed to characterize the sensor's heading error, coherent gain, and timebase stability. These measurements were conducted in a magnetically shielded room (MSR), located at the Technical University of Munich in Garching, to isolate the sensors from ambient magnetic noise.

\subsection{Heading Error and Direction Cosine Determination}
\label{app:heading_error}

A scalar magnetometer's reading can exhibit a systematic dependence on its orientation relative to the ambient magnetic field, an effect known as heading error \cite{Lee2021_HeadingErrors}. Minimizing and characterizing this error is crucial for accurately determining the background field direction $\hat{\mathbf{B}}_b$ and thus the projection cosines $(u_\theta, u_\phi)$ used in our signal model.

Our procedure involves mounting the sensor on a high-precision, non-magnetic rotation stage inside the MSR, where a stable, uniform DC magnetic field of known magnitude $|\mathbf{B}_{\text{cal}}| \approx \SI{48}{\micro\tesla}$ is applied. The sensor's orientation is then systematically varied by rotating it through a full $360^\circ$ range in both azimuth and tilt.

The sensor's output is recorded as a function of the rotation angles. The direction of the magnetic field is identified as the orientation that produces the maximum sensor reading. By comparing this experimentally determined direction to the nominal orientation set by the rotation stage, we can map out the heading error as a function of angle. For the main data taking, the sensors are aligned to this experimentally determined optimal direction to minimize the heading error. The residual uncertainty in this alignment, typically on the order of \SI{0.1}{\degree}, is the dominant contribution to the uncertainty in the direction cosines $(u_\theta, u_\phi)$ used in our systematic error budget (Sec.~\ref{subsec:systematic_budget}).

\subsection{Coherent Gain Calibration}
\label{app:gain_calibration}

The coherent gain determines the overall scaling of our signal model and thus directly impacts the final limit on $\varepsilon$. We calibrate the gain by injecting a weak, sinusoidal magnetic field of known amplitude and frequency and measuring the sensor's response.

Inside the MSR, a set of calibration coils is used to generate a small AC field, $\mathbf{B}_{\text{inj}}(t) = \mathbf{B}_0 \cos(2\pi f_{\text{inj}} t)$, superimposed on the large DC bias field. The frequency $f_{\text{inj}}$ is chosen to be within our primary search band (e.g., \SI{33}{\hertz}). The amplitude $B_0 = |\mathbf{B}_0|$ is precisely known based on the coil geometry and the current supplied by a calibrated function generator.

We record a long time series of the sensor's output and compute its spectrum. The coherent gain, $G(f_{\text{inj}})$, is then calculated as the ratio of the measured signal amplitude at $f_{\text{inj}}$ in the spectrum to the known input amplitude $B_0$. This procedure is repeated at several frequencies across our band of interest to verify that the gain is flat, as expected for these sensors in this frequency range. The uncertainty on the gain, estimated to be around 2\%, is dominated by the uncertainty in the calibration coil's geometric factor and the current source's amplitude accuracy. This 2\% uncertainty on the gain propagates linearly to a 2\% systematic uncertainty on our final limit for $\varepsilon$.

\subsection{Timebase Stability and Time Jump Characterization}
\label{app:timebase_stability}

A stable timebase is essential for maintaining signal coherence over the duration of an analysis segment. We characterized two aspects of the timebase: long-term frequency drift and short-term sample dropping events ("time jumps").

\paragraph{Long-Term Frequency Drift.}
To measure the long-term stability of the sensor's internal sampling clock, we used the same setup as for the gain calibration, injecting a highly stable \SI{33}{\hertz} tone for an extended period of over 18 hours. The data was divided into short, consecutive windows (e.g., 60 seconds), and a Lorentzian function was fitted to the peak in the spectrum of each window to determine its central frequency with high precision.

Plotting the fitted peak frequency over time reveals a very slow, linear drift. We measured this drift to be approximately \SI{73}{\nano\hertz\per\second}. This corresponds to a frequency change of only about \SI{0.26}{\milli\hertz} over a one-hour period, which is a tiny fraction of a frequency resolution bin ($1/T$). We therefore conclude that long-term clock drift does not significantly impact the coherence of the signal within a single analysis segment and can be safely neglected.

\paragraph{Time Jump Characterization.}
A critical instrumental artifact identified during our characterization is that the sensors are prone to sporadically dropping one or more samples without flagging an error. To characterize the rate and nature of these "time jumps," we analyzed long datasets by comparing the recorded data to a perfectly predictable, known input signal (such as the injected \SI{33}{\hertz} sine wave). A time jump manifests as a sudden phase shift in the residual between the measured data and the known signal.

By scanning the data for such phase discontinuities, we determined that these jumps occur at random intervals, with an average rate of approximately one event every 14 seconds in our test data, although the rate can vary. Each jump typically corresponds to a loss of a few samples. As discussed in Sec.~\ref{subsec:preprocessing}, our analysis pipeline is designed to detect and reject any data segment containing such a jump, thereby preserving the integrity of the coherent analysis at the cost of a small reduction in the total effective observation time.

\bibliographystyle{apsrev4-2}
\bibliography{refs2}           

\end{document}